\documentclass[useAMS,usenatbib]{mn2e}
\usepackage{graphicx}
\usepackage{textcomp}
\usepackage{txfonts}
\usepackage{comment}
\usepackage{capt-of}
\usepackage[hyperindex,breaklinks=true,colorlinks,citecolor=blue]{hyperref}
\usepackage{subfigure}
\usepackage{bm}
\usepackage[titletoc]{appendix}

\newcommand{\be}{\begin{equation}}
\newcommand{\ee}{\end{equation}}
\newcommand{\ba}{\begin{eqnarray}}
\newcommand{\ea}{\end{eqnarray}}

\newcommand{\nn}{\nonumber \\}

\newcommand{\rgl}{\rangle} 
\newcommand{\lgl}{\langle}

\newcommand{\lm}{{\ell m}}

\newcommand{\Ylm}{Y_\lm}
\newcommand{\sYlm}{\,_sY_\lm}

\newcommand{\tYlm}{\,_2\!Y_\lm} 
\newcommand{\mtYlm}{\,_{-2}\!Y_\lm}

\newcommand{\lsim}{\mathrel{\rlap{\lower4pt\hbox{\hskip1pt$\sim$}}
    \raise1pt\hbox{$<$}}}                
\newcommand{\gsim}{\mathrel{\rlap{\lower4pt\hbox{\hskip1pt$\sim$}}
    \raise1pt\hbox{$>$}}}
\newcommand{\vgamma}{\bmath{\gamma}}
\newcommand{\vtheta}{\bmath{\theta}}

\title[Galaxy-galaxy and galaxy-cluster lensing with the SDSS and
  VLA FIRST surveys]{Galaxy-galaxy and galaxy-cluster lensing with the
  SDSS and FIRST surveys} \author[Demetroullas \&
  Brown]{C.~Demetroullas$^{1}$\thanks{\href{mailto:constantinos.demetroullas@gmail.com}{\nolinkurl{constantinos.demetroullas@gmail.com}}}
  \&
  M.~L.~Brown$^{1}$\thanks{\href{mailto:m.l.brown@manchester.ac.uk}{\nolinkurl{m.l.brown@manchester.ac.uk}}}\\ $^{1}$Jodrell
  Bank Centre for Astrophysics, School of Physics and Astronomy, The
  University of Manchester, Manchester, M13 9PL, UK. \\ }

\begin{document}
\pdfpageheight 11.692in 
\date{Accepted 2015 XXXXX XX. Received 2015 XXXXX XX; in original form 2015 XXXXX XX}

\pagerange{\pageref{firstpage}--\pageref{lastpage}} \pubyear{2015}

\maketitle

\label{firstpage}

\begin{abstract}
We perform a galaxy-galaxy lensing study by correlating the shapes of
$\sim\!\!2.7\times10^5$ galaxies selected from the VLA FIRST radio
survey with the positions of $\sim$38.5 million SDSS galaxies,
$\sim$132000 BCGs and $\sim$78000 SDSS galaxies that are also detected in
the VLA FIRST survey. The measurements are conducted on angular
scales $\theta \lesssim 1200$ arcsec. On scales
$\theta \lesssim 200$ arcsec we find that the measurements are corrupted
by residual systematic effects associated with the instrumental beam
of the VLA data. Using simulations we show that we can successfully apply
a correction for these effects. 
Using the three lens samples (the SDSS DR10 sample, the BCG sample and
the SDSS-FIRST matched object sample) we measure a tangential shear
signal that is inconsistent with zero at the 10$\sigma$, 3.8$\sigma$
and 9$\sigma$ level respectively. Fitting an NFW model to the detected
signals we find that the ensemble mass profile of the BCG sample
agrees with the values in the literature. However, the mass profiles
of the SDSS DR10 and the SDSS-FIRST matched object samples are found
to be shallower and steeper than results in the literature
respectively. The best-fitting Virial masses for the SDSS DR10, BCG
and SDSS-FIRST matched samples, derived using an NFW model and
allowing for a varying concentration factor, are ${\rm M}_{200}^{\rm
  SDSS-DR10} = (1.2 \pm 0.4) \times10^{12} {\rm M}_{\odot}$, ${\rm
  M}_{200}^{\rm BCG} = (1.4 \pm 1.3) \times 10^{13} {\rm M}_{\odot}$
and ${\rm M}_{200}^{\rm SDSS-FIRST} =8.0 \pm 4.2 \times 10^{13} {\rm
  M}_{\odot}$ respectively. These results are in good agreement
(within $\sim2\sigma$) with values in the literature. Our findings
suggest that for galaxies to be both bright in the radio and in the
optical they must be embedded in very dense environment on scales $R
\lesssim 1$ Mpc.
\end{abstract}

\begin{keywords}
gravitational lensing: weak, methods: statistical, cosmological
parameters, galaxies: distances and redshifts
\end{keywords}

\section{Introduction}\label{sec:intro}
Galaxies can be categorised into two main types; the spiral star
forming late type and the elliptical passively evolving early type
ones. Furthermore, the spiral galaxies form a sequence defined by their
bulge size. Spirals with extremely large bulges are often grouped with
elliptical galaxies and are also referred to as early types. Early
type galaxies tend to be redder than late type galaxies
\citep{wong2012,tojeiro2013}. Observational and theoretical evidence
points to the importance of environment conditions on the properties
of galaxies. Early type galaxies for example are usually found in less
dense areas compared to late type galaxies
\citep[e.g.][]{dressler1980}. \citet{blanton2005b} found that galaxy
colours and luminosities correlate with the galaxy densities. Further,
star-formation is strongly associated with the density on small
($\sim$1Mpc) scales \citep{balogh2004,blanton2006}. Finally, measuring
dark matter profiles of galaxies can provide us with information
related to the galaxies' merging history \citep{mandelbaum2006b}.

Clusters of galaxies, on the other hand, are among the most promising
probes of cosmology and the physics of structure
formation. Theoretical predictions \citep{gunn1972,press1974} followed
by numerical simulations \citep{navarro1997,evrard2002} have shown
that rich clusters are associated with the most massive collapsed
haloes. N-body simulations have predicted that dark matter halos
should follow a Universal density profile
\citep{navarro1996,navarro1997,moore1999,fukushige2001}. The
simulations have shown that cluster-size halos should have a
relatively shallow and low-concentration mass profile with a density
that decreases with increasing radius. Moreover, studies have shown
that the evolution of cluster abundance with redshift is a function of
a number of cosmological parameters, in particular the normalisztion
of the matter power spectrum,  $\sigma_8$, and the dark energy equation
of state, $w$ \citep{white1993,viana1999,newman2002,bahcall2003}. 

An important parameter that needs to be studied in order to better
understand galaxies and galaxy clusters is the tidal gravitational
field that they reside in. This gravitational potential is generated
from the luminous and dark matter in the vicinity of the galaxy. The
visible component of the field may be extracted observationally using
galaxy properties such as luminosities or stellar mass. The dark
matter on the other hand can not be directly observed, but can be 
probed indirectly by its gravitational influence on its
surroundings. One such probe is the technique of gravitational lensing
which provides powerful measurements of the dark matter distribution in
the Universe across a wide range of scales (see
e.g. \citealt{massey2010} for a review).

Recent years have seen tremendous progress in the detection of weak
lensing by galaxies (hereafter galaxy-galaxy lensing) and by galaxy
clusters
\citep[e.g.][]{brainerd1996,mckay2001,hoekstra2003,sheldon2004,heymans2008,sheldon2009,velander2013}.
All of these experiments were conducted in the optical and/or near
infra-red (NIR) wavebands. The main reason for this (and in particular
for why radio surveys have not traditionally been used) is that at
current telescope sensitivity levels, significantly more galaxies are
visible in the optical/NIR sky compared to the radio sky. The deep,
wide field optical surveys typically used for weak lensing studies
routinely deliver $\sim 10$ galaxies arcmin$^{-2}$. In contrast, 
the deepest pencil-beam surveys performed in the radio band to date
only reach number densities of $\sim 1 - 2$ arcmin$^{-2}$, and even
then, this is typically achieved over only a very small sky area
\citep[e.g.][]{muxlow2005}. 

Nevertheless, using radio information in weak lensing studies can in
principle be very advantageous, and radio-based analyses are expected
to become competitive with the optical when future radio instruments,
such as the Square Kilometre Array, come online \citep{brown2015,
  harrison2016, bonaldi2016}. Radio interferometers have a well known
and deterministic beam pattern. The instrumental effects can therefore
be modelled and removed very accurately. In contrast, in optical weak
lensing studies the telescope point spread function (PSF) is known
precisely only at discrete locations on the sky, where a point source
has been detected, limiting its accurate decomposition.

Additional benefits arise when radio and optical data are
combined. Radio surveys are expected to be more sensitive to
populations of galaxies that are at higher redshifts compared to what
is typically probed in the optical. Combined studies can therefore
probe the Universe at earlier times. Additionally galaxy-galaxy
lensing depends on the lens-background object configuration, with the
signal strength weakened when such pairs are close in
redshift. Another advantage of combining optical and radio data for
weak lensing studies by galaxies, or by galaxy clusters, lies in the
fact that the two surveys trace source populations that are, in
general, well separated in redshift. This configuration therefore
helps to boost the signal and allows for a clearer identification of
the lens and source populations. Moreover as demonstrated in
\citet{demetroullas2015}, an optical-radio combined analysis can be
used to suppress position-correlated shape systematics in the data,
which, if uncorrected, can generate spurious signals that are orders
of magnitude stronger than the sought-after signal. This attribute is
particularly relevant for cosmic shear analyses which aim to measure
the minute weak lensing signal in background galaxies caused by the
intervening large-scale structure of the Universe
\citep[c.f.][]{harrison2016, camera2016}.

Finally, combining the data from radio and optical surveys allows one
to define novel galaxy lens samples, e.g.~jointly in terms of their
optical brightness, galaxy morphology, radio luminosity, AGN activity,
or gas abundance. We demonstrate this latter feature in this study by
defining one of our lens samples to be those galaxies that are bright
at both optical and radio wavelengths.

In this study, we use radio data from the Very Large Array (VLA) Faint
Images of the Radio Sky at Twenty centimetres (FIRST) survey
\citep{becker1995} in combination with optical data from the 10th Data
Release (DR10) of the Sloan Digital Sky Survey \citep{york2000, ahn2014} to
demonstrate the benefits of combining optical and radio surveys for
galaxy-galaxy and galaxy-cluster lensing analyses. By
cross-correlating the galaxy position and shape information from
the optical and radio data respectively, we study the ensemble mass
profiles of (i) a sample of optically-selected galaxies, (ii) a sample
of galaxy clusters and (iii) a sample of galaxies that have been found
to be bright in both wavelengths. 

The paper is organised as follows. In Section\,\ref{wlb} we present
the background weak lensing formalism and in Section\,\ref{dmhm} we
describe the dark matter halo models that we use later to interpret
our measurements. In Section\,\ref{Wldata} we describe the data-sets
that were used. Section\,\ref{FIRSTsyst} describes the shape
systematics that were identified in FIRST, and that are relevant for
this study, and the approach that was adopted to remove them. In
Section\,\ref{simsggl} we describe the simulations that we performed
to validate our systematics removal technique and to estimate the
uncertainties in our measurements from the real data. In
Section\,\ref{rdm} we present the results of the analysis of the real
data where we measure both galaxy-galaxy and galaxy-cluster
lensing signals. We discuss the results and conclude in Section\,\ref{con}.

\section{Weak Lensing Background}\label{wlb}
In the absence of intrinsic galaxy alignments, the weak lensing shear field can be
estimated by averaging over galaxy ellipticities:
\begin{equation}
\gamma_i = \lgl \epsilon_i \rgl,
\label{eq:ellip_expval}
\end{equation}
where $\gamma_1=\gamma \cos(2\phi)$ and $\gamma_2=\gamma
\sin(2\phi)$ are the two components of the spin-2 shear field,
$\vgamma = \gamma_1 + i \gamma_2$, with equivalent expressions for the
spin-2 ellipticity field. Here, $\phi$ is the angle
that the shear forms relative to an arbitrary reference axis, and the
brackets in equation~(\ref{eq:ellip_expval}) denote an ensemble
average. 

Expanding $\vgamma$ and its complex conjugate in terms of the
spin-weighted spherical harmonics, $\sYlm$ \citep{newman1966} as
\ba 
\vgamma (\Omega)  &=& \gamma_1 (\Omega) + i \gamma_2 (\Omega) \nn
	    &=& \sum_\lm (\kappa_\lm + i \beta_\lm)\tYlm(\Omega), \\
\vgamma^*(\Omega)  &=& \gamma_1 (\Omega) - i \gamma_2 (\Omega) \nn
	    &=& \sum_\lm (\kappa_\lm - i \beta_\lm)\mtYlm(\Omega), 
\ea 
where $s$ denotes the spin and the summation in $m$ is over $-\ell \le
m \le \ell$, one identifies the so-called $E$- and $B$-mode components
of the shear field with the lensing convergence field (denoted
$\kappa$) and the odd-parity divergence field (denoted
$\beta$). We note that the $\beta$ field is expected to be zero in
standard cosmological models and is therefore often used as a tracer
of systematic effects. The power spectrum of the convergence field, 
\be
C_\ell^{\kappa\kappa} = \frac{1}{2\ell+1} \sum_m  \kappa_\lm \, \kappa^*_\lm, \label{eq:clkk_est}\\
\ee
can be related to the 3-D matter power spectrum $P_{\delta}(k,r)$
through (e.g.~\citealt{bartelmann2001})
\begin{equation}
C_\ell^{\kappa\kappa}=\frac{9}{4} \left( \frac{H_0}{c} \right)^4
\Omega^2_m \int ^{r_H}_0 dr \, P_{\delta}\left(
\frac{\ell}{r},r\right)\left(\frac{\overline{W}(r)}{a(r)} \right)^2,
\label{eq:clkk_theory}
\end{equation}
where $H_0$ is the Hubble constant, $a$ is
the scale factor of the Universe, $r$ is comoving distance, $r_H$ is
the comoving distance to the horizon and $\Omega_m$ is the matter
density. The weighting, $\overline{W}(r)$, is given in terms of the
normalised comoving distance (or redshift) distribution of the source
galaxies, $G(r)dr = p(z)dz$:
\begin{equation}
\overline{W}(r)\equiv \int ^{r_H}_r dr' \, G(r')\frac{r' - r}{r'}.
\label{eq:W_of_r}
\end{equation}

In a similar fashion, one can expand the galaxy over-density field
$G$, at an angular position $\boldsymbol{\Omega}$, in terms of the
(spin-0) spherical harmonics $\Ylm$ as
\begin{equation}
G(\boldsymbol{\Omega})=\sum _{\ell m} G_{\ell m} \Ylm(\boldsymbol{\Omega}).
\end{equation}
The power spectrum of such an overdensity map ($C_\ell = 1 / (2\ell+1)
\sum_m G_{\ell m} G^*_{\ell m}$) can be related to the matter power
spectrum as
\begin{equation}
C_{\ell}^{GG}=\int ^{r_H}_0 \mathrm{d}r P_{\delta}\left(
\frac{\ell}{r},r\right) \frac{G(r)}{r} b^2_g\left(\frac{\ell}{r},r\right)~,
\label{eq:clgg_theory}
\end{equation}
where $G(r)$ is the normalized comoving distance distribution of the
galaxies in the over-density map and $b_g(k,r)$ is a scale- and
redshift-dependent bias describing the deviation of the galaxy
clustering from the dark matter clustering.

By correlating the values in an over-density map with the $\kappa$ and
$\beta$ fields of an overlapping shear map, one can construct the two
cross-power spectra,
\begin{equation}
C_{\ell}^{G\kappa} = \frac{1}{2\ell+1} \sum _m G_{\ell m}\kappa^*_{\ell m}~,
\end{equation}
\begin{equation}
C_{\ell}^{G\beta} =  \frac{1}{2\ell+1} \sum _m G_{\ell m}\beta^*_{\ell m}~.
\end{equation}
As mentioned above, the shear divergence field, $\beta$ is expected to be
zero and so we expect $C_\ell^{G\beta} = 0$. The galaxy over-density
convergence cross-spectrum can be related to the 3-D matter power
spectrum through \citep[e.g.][]{joachimi2010}
\begin{equation}
C_{\ell}^{G\kappa}=\frac{3}{2} \left( \frac{H_0}{\mathrm{c}}\right)^2
\Omega_{\mathrm{m}} \int ^{r_H}_0 \mathrm{d}r P_{\delta}\left(
\frac{\ell}{r},r\right)\left( \frac{\overline{W}(r) G(r)}{a(r) r}
\right) b_g\left(\frac{\ell}{r},r\right)
r_g\left(\frac{\ell}{r},r\right)~,
\label{overshearspeq}
\end{equation}
where once again $G(r)$ is the normalized comoving distance distribution of the
galaxies in the over-density map and $\overline{W}(r)$ is the
lensing weighting function appropriate for the sources in the shear map
(equation~\ref{eq:W_of_r}). $r_g(k, r)$ is a scale- and
redshift-dependent cross-correlation coefficient describing the
stochasticity of the relationship between the dark matter and galaxy clustering. 

\subsection{Galaxy-galaxy lensing}
The shear components of equation~(\ref{eq:ellip_expval}) are defined with
respect to an arbitrarily defined reference axis. However, in the case where
one is interested in the weak lensing distortion in a population of
background sources due to the presence of a known foreground object
(or ``lens''), it is more natural to consider the tangential and
rotated shear (or ellipticity), defined by
\ba
\gamma_t &=&  \gamma_1 \cos(2\omega) + \gamma_2 \sin(2\omega), \label{eq:gammat} \\
\gamma_r &=& -\gamma_1 \sin(2\omega) + \gamma_2 \cos(2\omega), \label{eq:gammar}
\ea
where $\omega$ is the position angle formed by moving counter
clockwise from the reference axis to the great circle connecting each
source-lens pair. The tangential shear, $\gamma_t$, describes
distortions in a tangential and/or radial direction with respect to
the lens position. The rotated shear, $\gamma_r$, describes
distortions in the orthogonal direction, at an angle $\pm \pi/4$ to
the vector pointing to the lens position.

The galaxy over-density convergence angular cross power spectrum
(equation~\ref{overshearspeq}) can be related to
the tangential shear $\gamma_t(\theta)$ (where $\theta$ is the angular
separation between lens and source) through
\citep[e.g.][]{hu2004}
\begin{equation}
\gamma_t(\theta)=\frac{1}{2\pi}\int \mathrm{d}\ell \, \ell
C_{\ell}^{G\kappa} J_2({\ell \theta})~,
\label{tangshearst}
\end{equation}
where $J_2(x)$ is a Bessel function. Since, in the absence of
systematics, one expects $C_l^{G\beta} = 0$, one therefore also
expects that the ``rotated shear''
\begin{equation}
\gamma_r(\theta)=\frac{1}{2\pi}\int \mathrm{d}\ell \, \ell
C_{\ell}^{G\beta} J_2({\ell \theta})~,
\end{equation}
should be consistent with zero and hence can be used to trace
systematics in the data.

In this study we estimate $\gamma_t(\theta)$ and $\gamma_r(\theta)$ by
stacking the measured shapes of VLA FIRST galaxies around the
positions of samples of lens galaxies selected from the SDSS catalogue
(see Section~\ref{Wldata}). To estimate the tangential and rotated
shear from the data we use the simple estimators:
\ba \widehat{\gamma_t}(\theta) &=& \frac{1}{N} \sum_{ls}
\epsilon_t^{(ls)}(\theta), \label{eq:gammat_est}
\\ \widehat{\gamma_r}(\theta) &=& \frac{1}{N} \sum_{ls}
\epsilon_r^{(ls)}(\theta), \label{eq:gammar_est} \ea
where the tangential and rotated ellipticity ($\epsilon_t$ and
$\epsilon_r$) are defined analogously to the shear quantities
(equations \ref{eq:gammat} \& \ref{eq:gammar}) and the summation is
over all lens-source ($ls$) pairs (total number $N$) separated by an
angle $\theta$. In practice, we further average the
$\widehat{\gamma_t}(\theta)$ and $\widehat{\gamma_r}(\theta)$
estimates within linear bins in angular separation.

\section{Dark Matter Halo Models}\label{dmhm}
The theoretical description presented in Section~\ref{wlb} is an
accurate model for the linear, and mildly non-linear, evolution of the
density perturbations in the Universe. However, on galaxy and galaxy
cluster scales the evolution of matter perturbations becomes highly
non-linear and the above description breaks down. On these scales, and
under the assumption that the lenses have a surface mass density that
is independent of the position angle with respect to the lens centre,
their density profiles can be predicted using axially symmetric
density profile models.

\subsection{SIS Model}
One of the most widely used axially symmetric dark matter halo models
is the singular isothermal sphere (hereafter SIS). The model can be
derived assuming that the matter content is an ideal gas in
equilibrium confined in a spherically symmetric gravitational
potential.

For an isothermal equation of state the gravitational potential
($\Phi$) is given by
\begin{equation}
\Phi =-\sigma_u^2
\ln\left(\frac{\rho}{\rho_o}\right)=-\sigma_u^2\ln\left(\rho'\right),
\label{SISphi}
\end{equation}
where $\sigma_u$ is the velocity dispersion of the gas, $\rho$ is the
density of the source, $\rho_o$ is the mean density of the Universe
and $\rho'$ is the fractional density given by $\rho'= \rho / \rho_o$.

For a given velocity dispersion $\sigma_u$, by applying the Virial
theorem one can calculate the radius $R_{200}$ and mass $M_{200}$ of
the dark matter halo at which its density is equal to 200 times the
critical density of the Universe, $\rho_c = 3 H^2 / 8 \pi G$ (where
$H$ is the Hubble parameter and $G$ is Newton's gravitational
constant). 

Rearranging equation~(\ref{SISphi}) with respect to the matter density
$\rho'$ we get
\begin{equation}
\rho'=\exp\left({\frac{-\Phi}{\sigma_u^2}}\right).
\end{equation}
Inserting this in the Poisson equation $\nabla^2 \Phi=-4\pi G \rho$ we find
\begin{equation}
-\sigma_u^2 \frac{1}{r^2}\frac{\mathrm{d}}{\mathrm{d}r}r^2
\frac{\mathrm{d}}{\mathrm{d}r}\left(\ln\rho' \right)=4\pi G \rho',
\label{sisder}
\end{equation}
where $r$ is the radial component of the spherical coordinate
system. Integrating equation~(\ref{sisder}) we find
\begin{equation}
\rho'(r)=\frac{\rho}{\rho_o}=\frac{\sigma_u^2}{2\pi G r^2}.
\label{rhosis}
\end{equation}
Using equation~(\ref{rhosis}), and by projecting the three-dimensional
density along the line of sight we obtain the corresponding surface
mass density, $\Sigma$:
\begin{equation}
\Sigma(b)=2\frac{\sigma_u^2}{2\pi G}\int_0^{\infty}
\frac{\mathrm{d}z}{b^2 + z^2}=\frac{\sigma_u^2}{2 G b},
\end{equation}
where $b$ is the transverse impact parameter (perpendicular to the
line of sight) and $z$ is the direction along the line of sight. 
The model has two pathological properties: it predicts a total source
mass that is infinite and a surface mass density that goes to infinity
as we move close to the centre of the object ($b=0$). Nevertheless it
has been used in many studies as it has been shown to be a good fit to data
for angular scales $\theta < 3$ arcmin \citep{vanuitert2011}.

The corresponding dimensionless surface mass density is 
\begin{equation}
\kappa(\boldsymbol{\theta})=\frac{\theta_E}{2\boldsymbol{\theta}},
\label{kappasis}
\end{equation}
where $\boldsymbol{\theta}$ is the projected 2-D angular position, and
where the Einstein radius $\theta_E$ is equal to 
\begin{equation}
\theta_E=4\pi\left(\frac{\sigma_u}{c}\right)^2 \frac{D_{LS}}{D_S}.
\end{equation}
Here $D_{LS}$ and $D_{S}$ are the lens-source distance and the
distance to the source respectively. In the SIS model the shear
$\boldsymbol{\gamma}$ relates to the Einstein radius as
\begin{equation}
\vgamma(\boldsymbol{\theta})=-\frac{\theta_E}{2|\boldsymbol{\theta}|}e^{2i\alpha},
\label{gammasis}
\end{equation}
where $\alpha$ is the polar angle of the galaxy position relative to
the lens centre. 
Equation~(\ref{gammasis}) shows that for an axially-symmetric mass
distribution, the shear is always tangentially aligned relative to the
mass centre. Expressing $\boldsymbol{\gamma}$ in terms of its
tangential ($E$-mode) and rotated ($B$-mode) components, for this
circularly symmetric lens model, we find (setting $\theta \equiv
|\boldsymbol{\theta}|$)
\begin{equation}
\gamma_t(\theta)=\frac{\theta_E}{2\theta}~~~~~ \mathrm{and}~~~~~~ \gamma_r(\theta)=0.
\label{gammasis2}
\end{equation}
That is, the shear field should only include a tangential $E$-mode component
and, as noted already, the rotated $B$-mode component can be used to
test for systematics in the data. 

\subsection{NFW Model}
\citet{navarro1996} (hereafter NFW) found, using simulations in the
framework of CDM cosmology, that the density profile of dark matter
halos for objects with masses in the range $10^{12} \lesssim
M/h^{-1}\mathrm{M}_{\odot} \lesssim 10^{15}$ can be accurately
represented by the radial function 
\begin{equation}
\rho(r)=\frac{\rho_c \delta_c}{(r/r_s)(1+r/r_s)^2},
\label{rhonfw}
\end{equation}
where the ``scale radius'' $r_s=r_{200}/c$ is the characteristic radius of
the object. $r_{200}$ is the radius of the object where its density is
equal to 200 times the critical density of the Universe $\rho_c$, $c$
is a dimensionless number known as the concentration parameter and
\begin{equation}
\delta_c = \frac{200}{3}\frac{c^3}{\ln(1+c)-c/(1+c)}.
\end{equation}
The mass of a NFW halo contained within a radius $r_{200}$ is therefore
\begin{equation}
M_{200}\equiv M(r_{200})=\frac{800\pi}{3}\rho_cr^3_{200}=\frac{800\pi}{3}\frac{\bar{\rho}(z)}{\Omega(z)}r^3_{200}~,
\end{equation}
where $\bar{\rho}(z)$ and $\Omega(z)$ are the mean density and the
density parameter of the Universe at redshift $z$.

Several algorithms have been developed to estimate the concentration
of dark matter halos. All are based on the assumption that the density
of the halos reflects the mean cosmic density at the time the halo had
formed. This is justified with simulations of structure formation
which showed that halos were more concentrated the earlier they were
formed. As expected, all models predict that the concentration factor
depends on cosmology. Additionally, all models predict that the
concentration increases towards lower masses, a direct result of less
massive systems collapsing at higher redshifts. For more details on
the various algorithms for calculating the concentration factors see 
\citet{Meneghetti} and references therein.

The logarithmic slope of the NFW density profile changes from $-1$ at
the centre of the object to $-3$ at large radii. The model therefore
predicts a mass density that is flatter than the SIS for the inner
part of the halo and steeper in the outskirts. Additionally, in
contrast to the SIS, the NFW model has no points where the mass density
$\rho$ becomes infinite, making it more realistic.

The NFW surface mass density is obtained by integrating the NFW mass
density profile (equation~\ref{rhonfw}) along the line of sight
\citep{wright2000}. Since the NFW is a spherically symmetric profile,
the radial dependence of the shear can be written as
\begin{equation}
\gamma_t^{\mathrm{NFW}}(x)=\frac{{\overline\Sigma^{\mathrm{NFW}}(x)}-\Sigma^{\mathrm{NFW}}(x)}{\Sigma_c}~,
\end{equation}
where we adopt the dimensionless radial distance $x=r/r_s$ and the
mean surface mass density ${\overline\Sigma^{\mathrm{NFW}}(x)}$ is
equal to
\begin{equation}
{\overline\Sigma^{\mathrm{NFW}}(x)}= \frac{2}{x^2}\int ^x _ 0  x'
\Sigma^{\mathrm{NFW}}(x') {\mathrm d}x'.
\end{equation}
The radial dependence of the shear is therefore \citep{wright2000}
{\large
\begin{equation}
\gamma_t^{\mathrm{NFW}}(x) = \left\{ \begin{array}{l l}
	\frac{r_s\delta_c\rho_c}{\Sigma_c}g_<(x), & x < 1 \\
	\frac{r_s\delta_c\rho_c}{\Sigma_c}\left[\frac{10}{3}+4\ln\left(\frac{1}{2}\right)\right]~,  & x = 1 \\
	\frac{r_s\delta_c\rho_c}{\Sigma_c}g_>(x), & x > 1~, \end{array} \right.
\end{equation}}
\noindent where the functions $g_<(x)$ and $g_>(x)$ are given by 
\begin{eqnarray}
g_<(x)&=&\frac{8\,\mathrm{arctanh}\sqrt{(1-x)/(1+x)}}{x^2\sqrt{x^2-1}}+ \frac{4}{x^2}{\mathrm \ln}\left( \frac{x}{2} \right)
-\frac{2}{(x^2-1)} \nn
&+&\frac{4\,\mathrm{arctanh}\sqrt{(1-x)/(1+x)}}{(x^2-1)\sqrt{1-x^2}}~, \\
g_>(x)&=&\frac{8\,\mathrm{arctan}\sqrt{(x-1)/(1+x)}}{x^2\sqrt{1-x^2}}+ \frac{4}{x^2}{\mathrm \ln}\left( \frac{x}{2} \right)
-\frac{2}{(x^2-1)} \nn
&+&\frac{4\,\mathrm{arctan}\sqrt{(x-1)/(1+x)}}{(x^2-1)^{3/2}}~.
\end{eqnarray}
In this study, in addition to directly fitting for the concentration factor
$c$ using our measurements from FIRST and SDSS, we also investigate
fixing the concentration parameter according to \citet{bullock2001}
who derived best-fitting mass- and redshift-dependent halo
concentration parameters from high-resolution $N$-body simulations of
a $\Lambda$CDM cosmology. 

\section{The Data}\label{Wldata}
The background objects used in this study are extracted from the VLA
FIRST catalogue. The sample contains information about $\sim$1 million
sources out of which $\sim$270,000 are resolved and can be used for
galaxy-galaxy lensing studies. Based on the simulations of
\cite{wilman2008} we have estimated the median redshift of the survey
to be $z^{\rm{FIRST}}_{\rm{median}}$$\simeq$1.2. We perform our
galaxy-galaxy lensing analysis using three different lens samples. The
first is drawn from the SDSS DR10 sample. The catalogue contains
$\sim$38.5 million entries and has a median redshift of
$z^{\rm{SDSS}}_{\rm{median}}$$\simeq$0.53 \citep{sypniewski2014}. For
more information about these two galaxy samples see Section~3 of
\citet{demetroullas2015}. The other two lens samples are described in
the following sub-sections.

\subsection{Brightest Cluster Galaxy Sample}

Gravitational lensing directly traces the matter distribution of the
deflecting object. It is therefore expected that the phenomenon will be
more apparent around massive objects like galaxy clusters. Brightest
Cluster Galaxies (BCGs) are luminous elliptical galaxies located at
the potential centres of clusters.  By using such objects as the lens
sample, one can therefore statistically examine the weak lensing
signal around galaxy clusters. We draw the positions for a number of these
objects from the BCG-based galaxy cluster catalogue of
\citet{wen2012}. That work used the SDSS Data Release 8 (DR8) to identify 132,684 BCGs
in the redshift range $0.05 < z < 0.8$. To identify a cluster, they
used the following criteria:
\begin{itemize}
\item The richness $R_{L\star}= L_{\mathrm{total}}/L_{\star} \geq 12$.
\item The number of galaxy candidates within a photometric redshift bin of 
$z \pm 0.04(1+z)$ and a radius $r_{200}$ should be $N_{200} \geq 8$. 
\end{itemize}
Here, $L_{\mathrm{total}}$ is the total luminosity of the member
galaxies in the r band and the characteristic luminosity of galaxies in
that band, $L_{\star}(z)=L_{\star}(z=0)10^{0.648z}$
\citep{blanton2003}. The brightest member within a radius of 0.5\,Mpc
from where the number density peaks is considered as the BCG.

The catalogue contains information about the cluster position, the
assigned photometric redshift $z_{\mathrm{ph}}$ and the r band
magnitude $r_{\mathrm{mag}}$. It also contains the radius and richness
of the cluster within the area in which its density is $\rho \geq
200\,\rho_{\mathrm{crit}} $. The median redshift for the sources in
the catalogue was calculated to be
$z_{\mathrm{\rm{median}}}^{\rm{BCG}}=0.37$.
  

\subsection{SDSS-FIRST Matched Object Sample}
The third lens sample is defined to be those galaxies that are visible
in both the radio and optical bands. We create a catalogue with SDSS
objects whose position matches the position of a FIRST source within a
5$''$ radius, which is the full-width at half maximum (FWHM) of the
VLA FIRST beam. The catalogue contains the combined information from
the FIRST and SDSS surveys for $\sim78,000$ galaxies. This 
population has a median redshift similar to the complete SDSS DR10
sample of $z_{\rm{median}}^{\rm{SDSS-FIRST}}$=0.57.

\section{FIRST shape corrections}\label{FIRSTsyst}
A detailed analysis of the shape systematics present in the FIRST data
was conducted by \citet{chang2004}. To minimize systematics that study
carefully selected those sources in the catalogue that were least
likely to be corrupted by telescope systematics. Following
\citet{demetroullas2015}, we apply as many of the \citet{chang2004}
selection criteria as possible, given the information that is
available in the FIRST catalogue. We therefore select only those
sources that are resolved and that have a deconvolved major axis size
in the range $2'' \leq \theta_{\rm maj} \leq 7''$. We also discard any
sources that have an integrated flux density $S_{1.4 {\rm GHz}} < 1$
mJy. Finally we remove all sources that are found to have a
possibility of being a sidelobe of a nearby brighter source that is
$P(S) \geq 0.05$. After applying these selection criteria to the data
we are left with $\sim 2.7\times10^{5}$ sources.

We then performed a quality assesment of the remaining FIRST sources,
the results of which showed that the data contain spurious
contributions to the source ellipticities
originating from an imperfect deconvolution and/or {\sevensize
  CLEAN}ing\footnote{%
{\sevensize CLEAN} \citep{hogbom1974} is a standard
technique commonly used in radio astronomy to deconvolve images for
the effects of a finite PSF. For a discussion of the effects of
{\sevensize CLEAN} in the context of galaxy shape measurements, see
\cite{tunbridge2016}.}
of the data. To assess the quality of the FIRST shape estimates, we
stacked the shapes of the $\sim 2.7\times10^{5}$ selected FIRST
sources around the positions of \emph{all} FIRST sources in the
catalogue (both resolved and unresolved). That is, we applied the
estimators of equations~(\ref{eq:gammat_est}) \& (\ref{eq:gammar_est})
using all of the FIRST radio sources as the lens sample. While these
constructions will contain some lens-source pairs which would be
expected to give rise to a genuine lensing signal, the vast majority
of the ``lenses'' in this sample are high-redshift ($z \sim 1$)
objects and so we expect the contribution from lensing to the overall
measured signal to be small.

\begin{figure*}
\subfigure{\includegraphics[width=5.5cm]{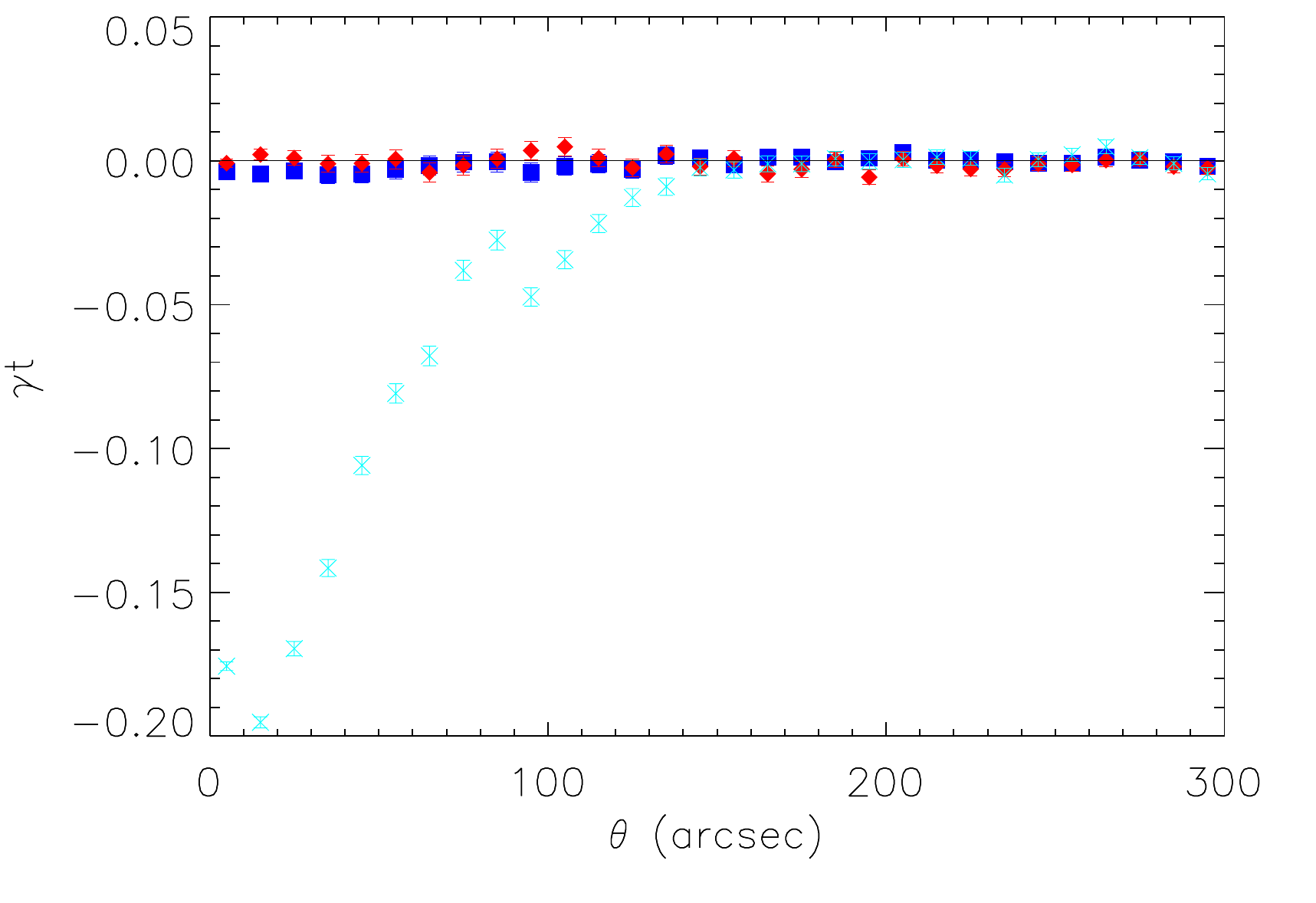}}\hspace{0.5em}
\subfigure{\includegraphics[width=5.5cm]{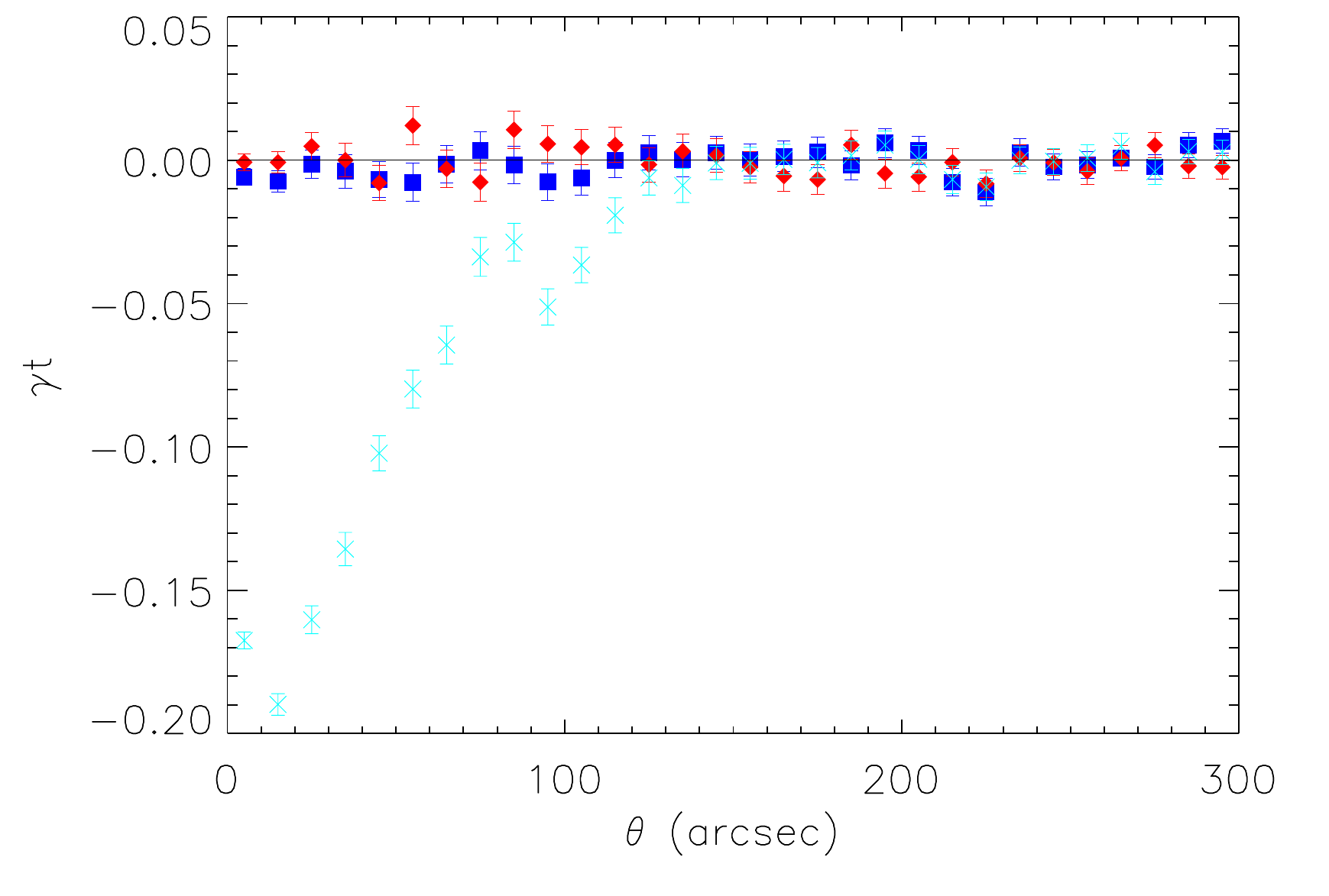}}\hspace{0.5em}
\subfigure{\includegraphics[width=5.5cm]{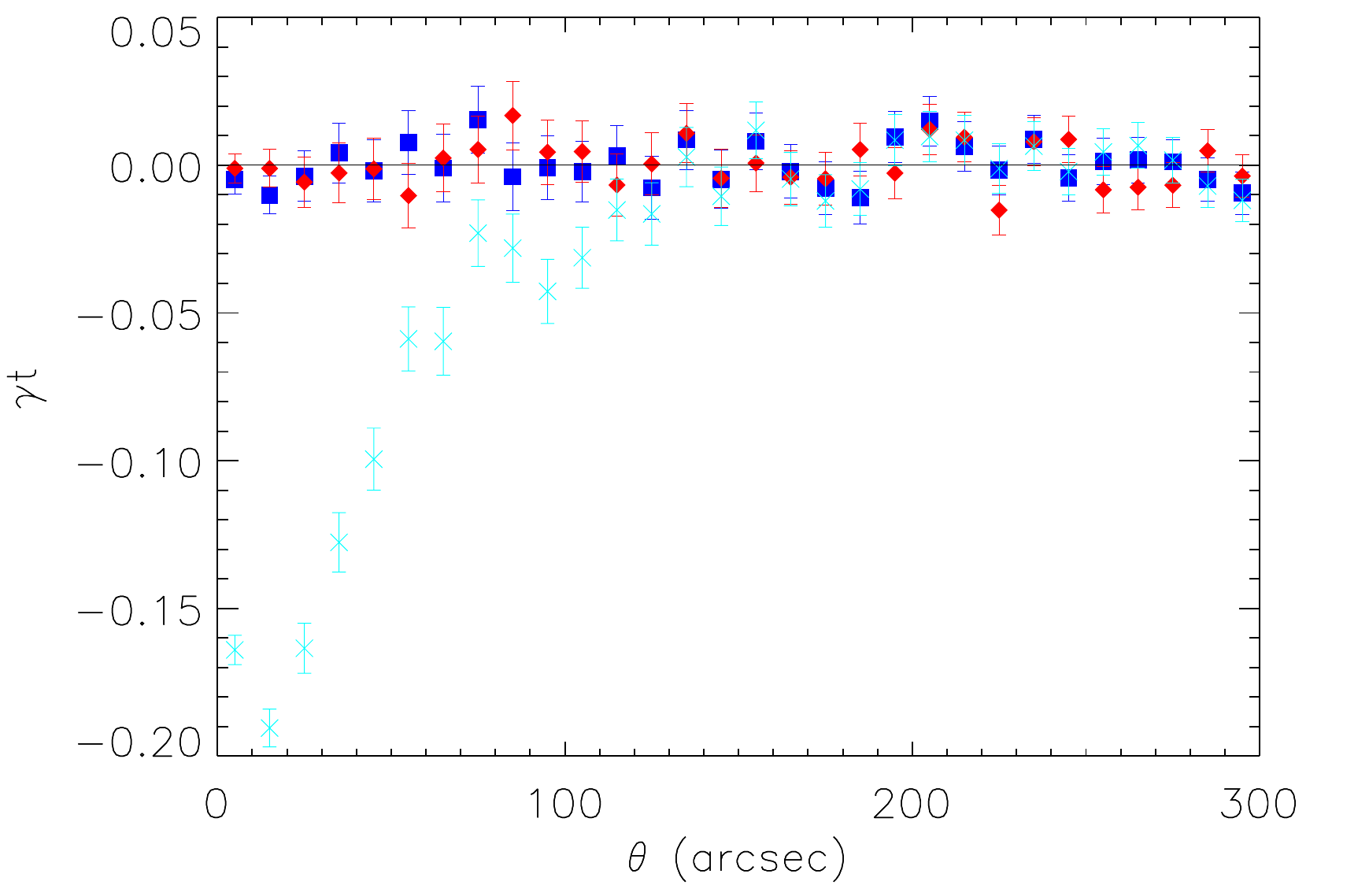}}
\caption{Tests for spurious distortions in the measured ellipticities
  of the FIRST background sources. The tangential shear, estimated
  using equation~(\ref{eq:gammat_est}), for a ``lens'' sample
  consisting of radio sources in the FIRST catalogue, is shown as
  the cyan crosses. A large radial systematic signal is apparent. From
  left to right, the three panels show the results when all the FIRST
  radio sources are used as the ``lens'' sample (\emph{left}), and randomly
  selecting 25\,\% (\emph{centre}) or 10\,\% (\emph{right}) of the FIRST sources
  respectively. The residual tangential and rotated shear after
  applying the correction described in Section~\ref{FIRSTsyst} are shown as the
  blue squares and red circles respectively.}
 \label{syst_FIRSTrandompick_FIRST005sidelobe_alltog}
\end{figure*}

The measurements of the tangential and rotated distortion resulting
from these tests are shown in the left panel of
Fig.~\ref{syst_FIRSTrandompick_FIRST005sidelobe_alltog} and reveal a
strong \emph{negative} tangential (i.e. radial) distortion signal 
$\widehat\gamma_t(\theta)$, on scales $\theta \lesssim 200''$. 
The measured rotated signal $\widehat\gamma_r(\theta)$, is consistent
with zero. Though noisier, the radial distortion signal persists if we
randomly select a subset of the FIRST objects as the ``lens'' sample
(middle and right panels of
Fig.~\ref{syst_FIRSTrandompick_FIRST005sidelobe_alltog}). Furthermore
we have confirmed that the measured signal shows no dependancy on the
flux or size of the objects used to construct either the lens or
source samples.

\begin{figure*}
\subfigure{\includegraphics[width=8.5cm]{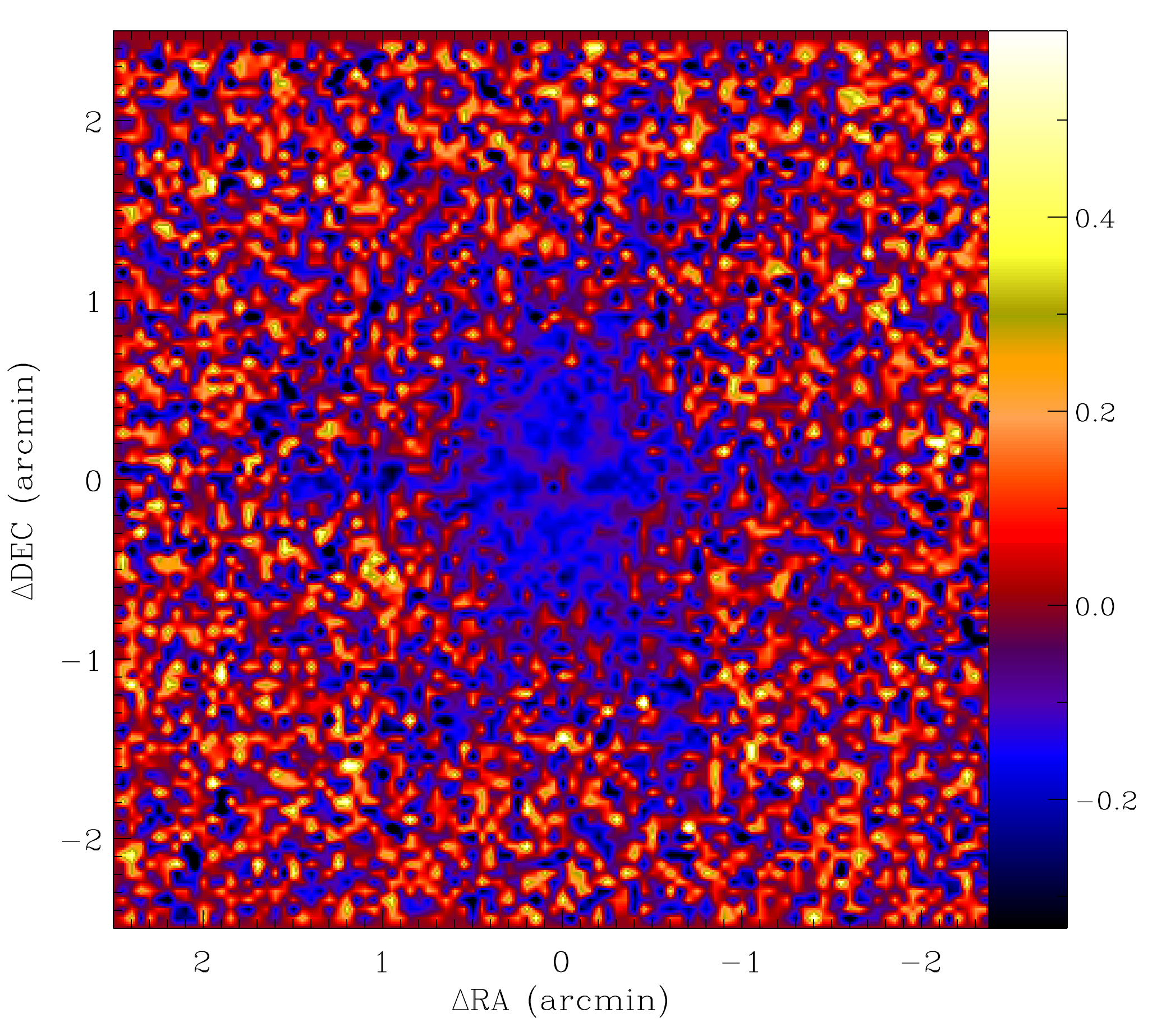}}\hspace{0.5em}
\subfigure{\includegraphics[width=8.5cm]{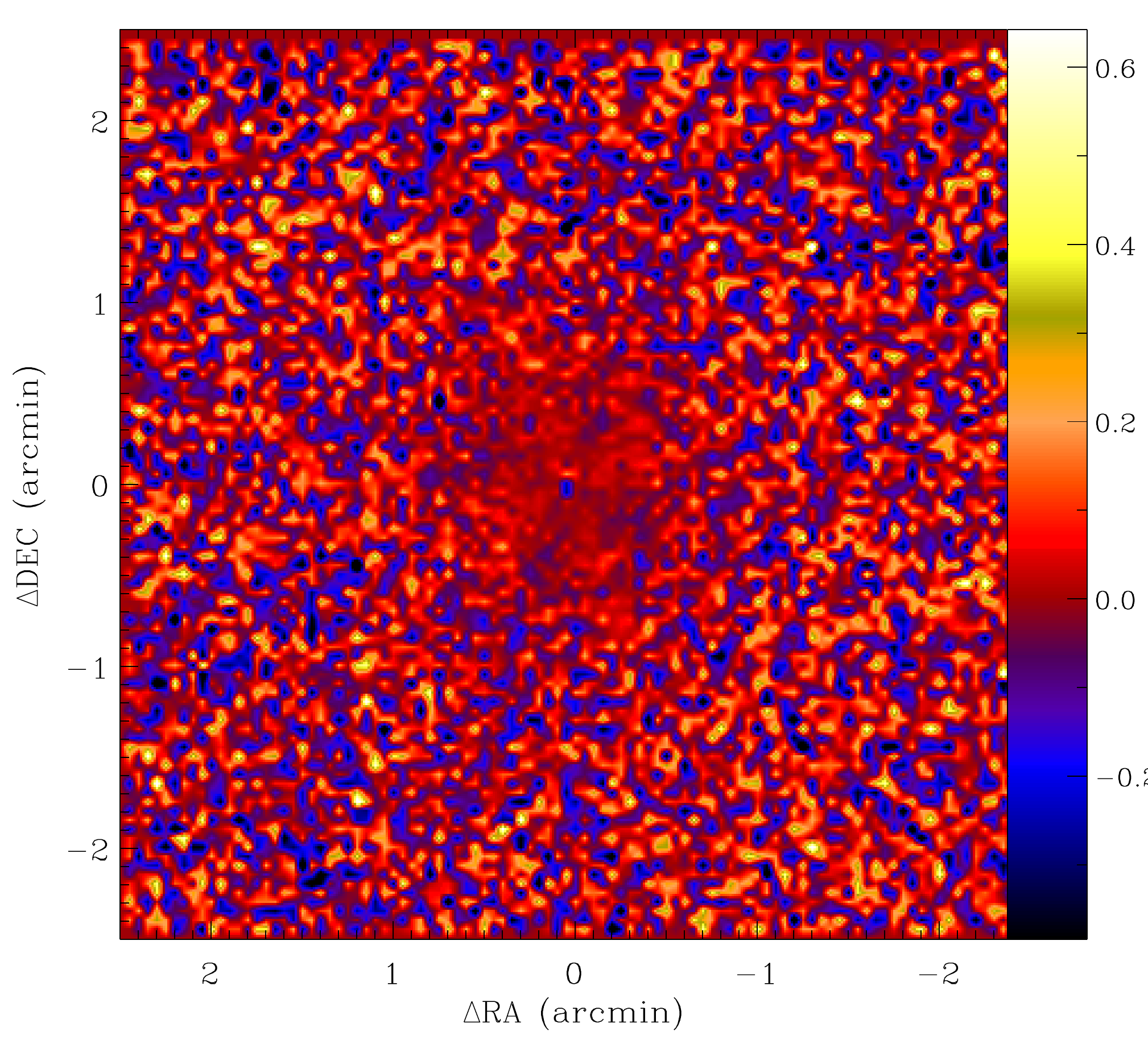}}
\subfigure{\includegraphics[width=8.5cm]{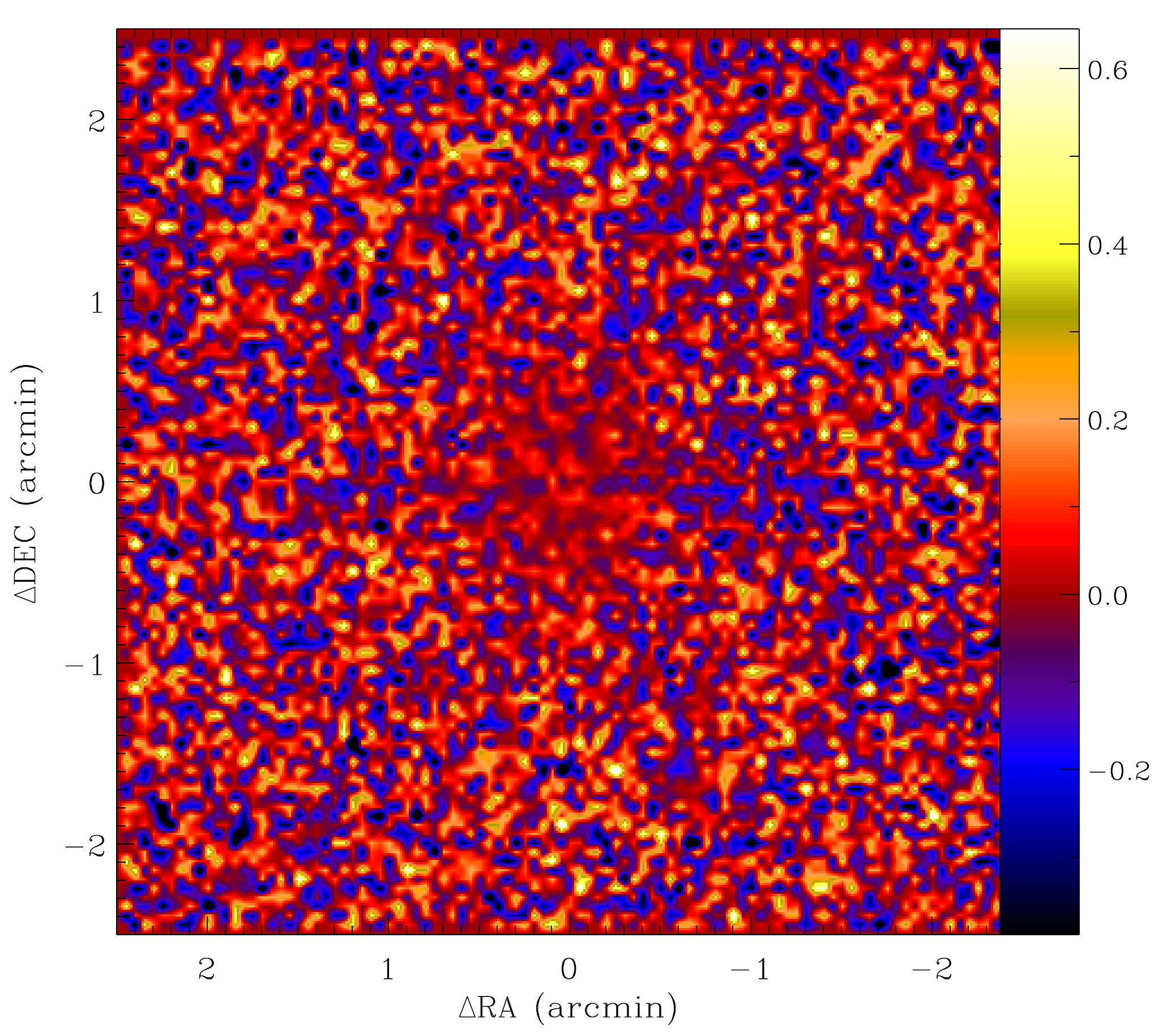}}\hspace{0.5em}
\subfigure{\includegraphics[width=8.5cm]{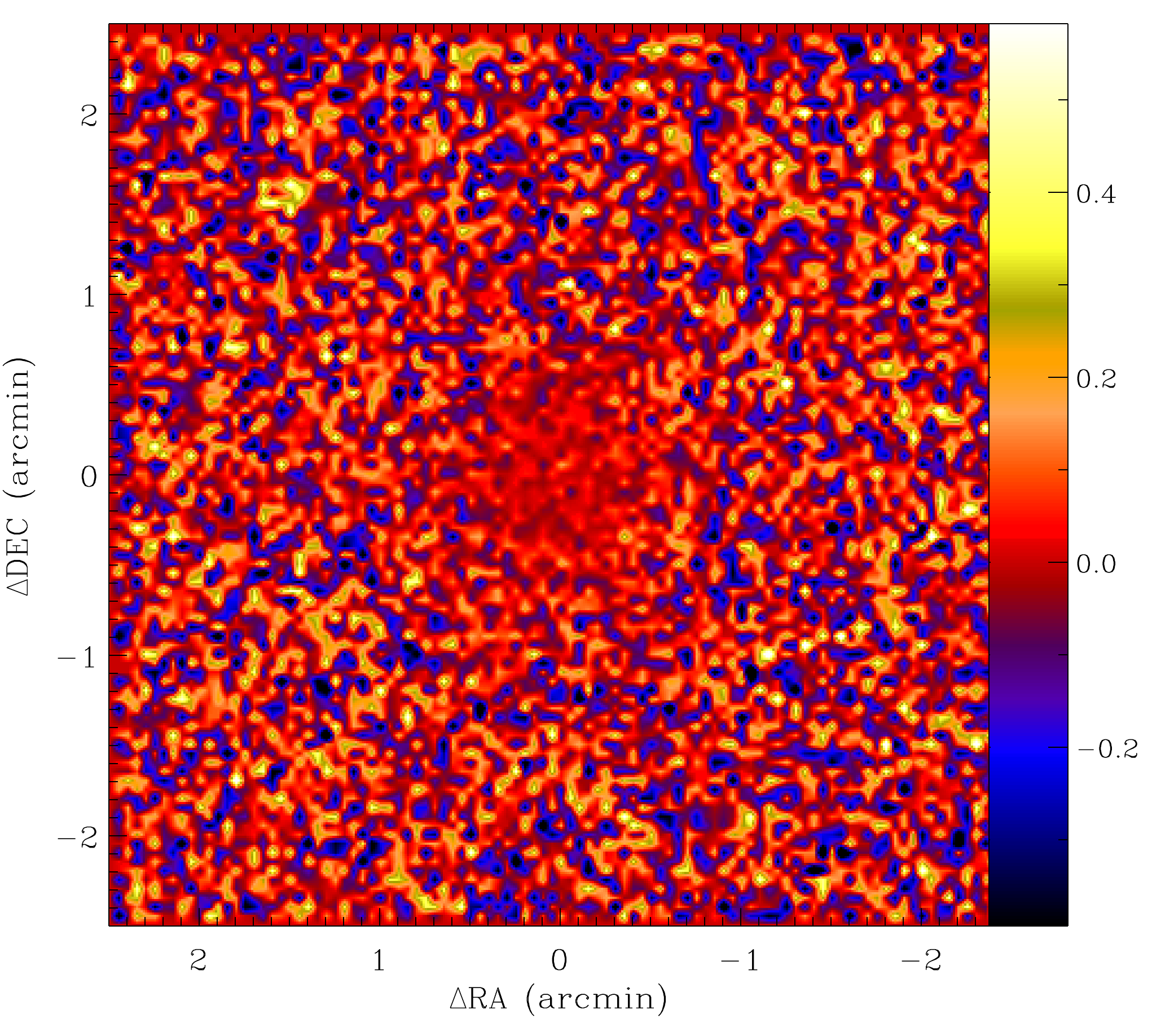}}
 \caption{Maps of the spurious tangential shear ($\gamma_t(\vtheta)$, \emph{left
     panel}) and rotated shear ($\gamma_r(\vtheta)$, \emph{right panel}) as a
   function of the 2-D displacement of the FIRST selected
   sources from central lens positions for the case where the ``lens''
   sample is composed of all sources in the FIRST catalogue.
   The \emph{upper} (\emph{lower}) panels show the measured
   distortion before (after) application of the correction described in
   Section~\ref{FIRSTsyst} respectively.} 
\label{fig:contdraddeccorr}
\end{figure*}

To investigate further, Fig.~\ref{fig:contdraddeccorr} shows the
tangential and rotated distortion measured from the same sample as
used in Fig.~\ref{syst_FIRSTrandompick_FIRST005sidelobe_alltog} but
now plotted as $\gamma_t(\vtheta)$ and $\gamma_r(\vtheta)$ where
$\vtheta = (\Delta RA, \Delta\delta)$ is the 2-D displacement vector
from the central lens positions.
This figure shows that the measured radial distortion signal resembles
the 6-arm shape of the synthesised beam (or PSF) of the VLA telescope
in ``snap-shot mode'' which was the observation mode that was employed
for the collection of the FIRST data. Once again the rotated shear
signal is found to be consistent with zero. The results clearly show that the
signal is of artificial origin and therefore the shapes need further
correction. The results also suggest that the measured systematic is
associated with an imperfect deconvolution and/or CLEANing of the
sources during the imaging stage of the FIRST data reduction.

These detected ellipticity distortions would most likely bias a
galaxy-galaxy or galaxy-cluster lensing analysis. We therefore apply a
correction to the shapes of the FIRST galaxies based on the spurious
negative tangential shear signal detected in
Fig.~\ref{syst_FIRSTrandompick_FIRST005sidelobe_alltog}. To apply the
correction, we assume that the detected $\gamma_t(\theta)$ signal in
the shape of each galaxy is due to an additive instrumental systematic
effect caused by the presence of all of the other sources in the FIRST
data (this would be the case if the origin of the systematic were an imperfect
deconvolution of the FIRST synthesised beam). In that case, for the
$i^{th}$ galaxy in the sample of FIRST background objects, we can
estimate the total contribution to its measured ellipticity from
instrumental systematic effects as
\ba
\epsilon_{1,i}^{\mathrm{sys}} &\approx& \sum_j
\gamma_t^{\mathrm{sys}}(\theta_{ij}) \cos(2 \omega_{ij}), \label{e1sp} \\ 
\epsilon_{2,i}^{\mathrm{sys}} &\approx& \sum_j
\gamma_t^{\mathrm{sys}}(\theta_{ij}) \sin(2 \omega_{ij}), \label{e2sp}
\ea
where the sum is over all sources in the FIRST
catalogue. $\omega_{ij}$ is the angle between the reference axis of
the $\epsilon_{1/2}$ coordinate system and the line joining the
$i^{th}$ galaxy in the FIRST background sample with the $j^{th}$
galaxy in the FIRST catalogue. $\gamma_t^{\rm sys}(\theta_{ij})$ is
the systematic tangential shear (plotted as the cyan points in the
left-hand panel of
Fig.~\ref{syst_FIRSTrandompick_FIRST005sidelobe_alltog}) and
$\theta_{ij}$ is the angular separation between the $i^{th}$ galaxy in
the FIRST background sample and the $j^{th}$ galaxy in the FIRST
catalogue. The FIRST shapes are then corrected simply by subtracting
the systematic signals, estimated using equations~(\ref{e1sp}) \&
(\ref{e2sp}), from the uncorrected galaxy ellipticities as listed in
the FIRST catalogue. 

The success of the correction algorithm was assessed by repeating the
diagnostic tests described above but now using the corrected
ellipticities. The results, shown in
Fig.~\ref{syst_FIRSTrandompick_FIRST005sidelobe_alltog} and
Fig.~\ref{fig:contdraddeccorr}, indicate that the correction has
successfully removed the systematic radial distortions that were
previously present in the FIRST shapes. 

The approach that we have used to estimate the 
instrumental systematics will be sensitive, to a small extent, to any
real galaxy-galaxy lensing signal present in the data.  This is
because the FIRST source population is extended in redshift space and
therefore some of the sources will be lensed by others in the
sample. We note that if this effect is present at a significant level,
then it is likely to dilute the radial distortion caused by the
instrumental systematics, thus leading to an under-correction of the
systematic effects. 
 
\section{Simulations}\label{simsggl}
As mentioned above, our procedure for correcting the FIRST galaxy
shapes for residual instrumental distortions will be sensitive, to a
small extent, to any real galaxy-galaxy lensing signal present in the
data. In this section, we use simulations to assess the extent of the
resulting bias in a subsequent galaxy-galaxy lensing study based on
the corrected shapes. The simulations are also used to further examine
the contamination removal method that was applied to the FIRST data
and to estimate the uncertainties in the measurement due to random
shape noise in the FIRST galaxies.

In order to generate the signal component of our simulations, we
follow the general procedure described in Section\,5 of
\cite{demetroullas2015} and model both the overdensity and shear signals as
correlated Gaussian random fields. Denoting overdensity fields
as traced by the FIRST and SDSS galaxy positions as
$G_f$ and $G_s$ respectively, and the convergence field extracted from
the FIRST shear measurements as $\kappa_f$, we can model the 2-point
statistics of these fields using the following power spectrum matrix:
\be
\mathbf{C}_\ell = \left(
\begin{array}{ccc} 
C^{G_fG_f}_\ell     & C^{G_f\kappa_f}_\ell     & C^{G_fG_s}_\ell \\
C^{\kappa_fG_f}_\ell & C^{\kappa_f\kappa_f}_\ell & C^{\kappa_fG_s}_\ell \\
C^{G_sG_f}_\ell     & C^{G_s\kappa_f}_\ell      & C^{G_sG_s}_\ell
\end{array}
\right).
\label{eq:cl_matrix}
\ee
Note that we do not need to include correlations involving the
convergence field extracted from SDSS shear measurements ($\kappa_S$)
in equation~(\ref{eq:cl_matrix}) as in this study we are not making
use of SDSS galaxy shapes. We then make the further simplifying assumption
that the overdensity field as traced by the FIRST galaxy positions is
not correlated with either of the other fields, in which case the
power spectrum matrix becomes:
\be
\mathbf{C}_\ell = \left(
\begin{array}{ccc} 
C^{G_fG_f}_\ell     & 0   &  0 \\
0 & C^{\kappa_f\kappa_f}_\ell & C^{\kappa_fG_s}_\ell \\
0     & C^{G_s\kappa_f}_\ell      & C^{G_sG_s}_\ell
\end{array}
\right).
\label{eq:cl_matrix_simp}
\ee
In reality there will be some correlation between the FIRST and SDSS
galaxy positions ($C^{G_fG_s}_\ell \equiv C_\ell^{G_sG_f} \neq 0$) since
the redshift distributions of these two samples do overlap (and indeed
there are $\sim78000$ galaxies common to both catalogues). Similarly,
there could also be a small correlation between the FIRST galaxy
positions and the convergence field as extracted from the FIRST shear
measurements ($C^{G_f\kappa_f}_\ell \equiv C_\ell^{\kappa_fG_f} \neq 0$)
due to lensing of background FIRST sources by foreground FIRST
sources. However, we expect both of these terms to be small and, in
any case, we are including the most important cross-correlation for
the purposes of the current study --- that between the convergence field as extracted from the FIRST
shear measurements and the SDSS galaxy positions ($C_\ell^{G_s\kappa_f}
\equiv C_\ell^{\kappa_f G_s}$). It is this latter correlation that gives
rise to a galaxy-galaxy (or cluster-galaxy) tangential lensing signal
that is measurable by stacking the shapes of background FIRST galaxies
around the positions of SDSS foreground objects.

We generate the entries of the power spectrum matrix
(equation~\ref{eq:cl_matrix_simp}) based on a $\Lambda$CDM cosmology
using equations~(\ref{eq:clkk_theory}), (\ref{eq:clgg_theory}) and
(\ref{overshearspeq}). For these calculations, we adopt the same
cosmological parameters as listed in Section\,5 of
\citet{demetroullas2015} and we assume median redshifts of
$z^{\rm{SDSS}}_{\rm{m}}=0.53$ and $z^{\rm{FIRST}}_{\rm{m}}=1.2$ for
the SDSS and FIRST samples respectively. (See Fig.\,8 in
\citealt{demetroullas2015} for the adopted redshift
distributions). For all of our simulations we assume that the galaxy
distribution traces the dark matter distribution perfectly
($r_g$=$b_g$=1). Once generated the power spectrum matrix is then used
to produce correlated realisations of the shear and over-density
fields within the {\sc healpix} framework \citep{gorski2005}
for the two surveys following the procedure outlined in Section\,5.1
of \cite{demetroullas2015}.

In order to probe the angular scales on which the residual distortions
in the FIRST data have been detected (see
Fig.~\ref{syst_FIRSTrandompick_FIRST005sidelobe_alltog}), the
simulated maps are generated at a {\sc healpix} resolution of {\sc
  nside} = 8192, corresponding to a sky pixel side length $\simeq\,25$
arcsec. 
The simulated over-density fields for the two surveys are used to
assign mock positions to 38.5 million SDSS and $2.7\times10^5$ FIRST
sources. Note that this analysis, contrary to
\citet{demetroullas2015}, makes use of only the positions of the SDSS
sources and not their shapes. Therefore all 38.5 million sources are
included in the simulation. 

For the FIRST survey we also generate a mock galaxy shape
catalogue as follows. Each FIRST simulated source is assigned
ellipticity components based on the values, at the appropriate sky
location, in the corresponding simulated shear map. Intrinsic shape
noise is included by adding a real FIRST galaxy ellipticity
measurement, randomly selected from the FIRST catalogue. The
systematic errors in FIRST induced by the VLA beam residuals are
modelled as described in Section\,5.3 of
\citet{demetroullas2015}. Briefly, we first generate a template of the
spurious negative tangential shear signal that was measured in the
FIRST data (Fig.~9 in \citealt{demetroullas2015}). The template is
normalised such that the amplitude of its azimuthicaly averaged
tangential shear signal matches the amplitude of the signal shown in
the left panel of
Fig.\,\ref{syst_FIRSTrandompick_FIRST005sidelobe_alltog}. The
normalized template is then used to generate additional spurious
ellipticity contributions which we add to the entries in our simulated
mock shape catalogue. For any one galaxy, these additional
contributions model the contamination due to the sidelobes of all of
the other sources in the data and are constructed from the normalized
template according to equations~(\ref{e1sp})--(\ref{e2sp}).

We create 100 simulated data-sets according to the above prescription, each one
containing a known galaxy-galaxy lensing signal and random ellipticity
noise due to the intrinsic shapes of the FIRST sources. In each case
we choose to store the information on the FIRST shapes prior to and
after the addition of our model for the instrumental
contamination. This allows us to assess the probable impact
of the residual beam effects on the measured galaxy-galaxy lensing
signal.

\subsection{Simulated Source Shape Corrections}
\begin{figure} 
\centering
\includegraphics[width=8.5cm]{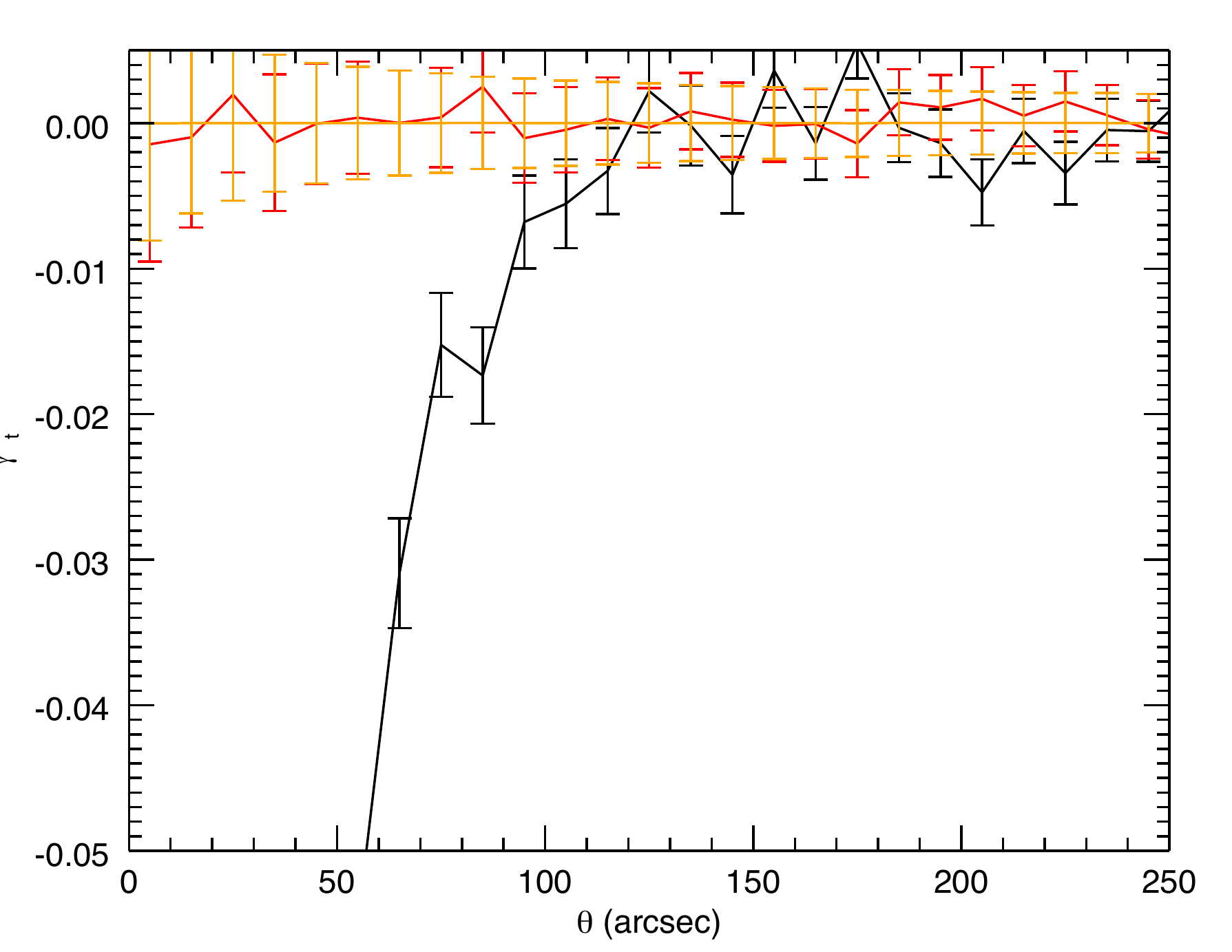} 
\caption{The measured contamination from a random simulation. The
  black, orange and red solid lines represent the measured tangential
  signal before (black points) and after (orange points) the simulated
  FIRST sources shapes were corrected and the measured rotated shear
  after the shape correction (red points).}
\label{fig:simcontcorr}
\end{figure}

Having created mock galaxy position and shear catalogues which include
a known galaxy-galaxy lensing signal, we then process the simulated
data in exactly the same way as used for the analysis of the real data. The
first step is to correct the mock FIRST galaxy shapes for the effects
of the spurious radial distortion signal that was included in our
simulations. We perform this correction as described in
Section~\ref{FIRSTsyst}, constructing an estimate of the systematic
signal by stacking the shapes of the FIRST sources around
the positions of the FIRST sources. The resulting template is then
used to correct the entries in the mock shape catalogues for the
systematic effect. A demonstration of this correction at work on
one of our simulations is presented in Fig.\,\ref{fig:simcontcorr},
which shows a radial systematic signal similar to that
found in the real data
(Fig.\,\ref{syst_FIRSTrandompick_FIRST005sidelobe_alltog}) being
successfully removed by our correction algorithm. 

\subsection{Galaxy-Galaxy Lensing Signal from Simulations}
We proceed to measure the galaxy-galaxy lensing signal by correlating
the positions of the SDSS objects with the shapes of the FIRST sources
in the mock catalogues. Before doing so however, we point out a number
of modifications that need to be applied to the theoretical prediction
for a self-consistent analysis.  

Firstly, we note that by imprinting a known theoretical
$C_\ell^{G\kappa}$ signal onto a pixelised map, its shape is
altered. To account for the effect of the pixelisation, we must
include its smoothing effect on the input power spectrum; that is, we
consider the effective input power to be $F_{\ell} ^{2}
C_{\ell}^{G\kappa}$, where $F_{\ell}$ is a known function describing
the smoothing effects of the pixelisation used in the {\sc healpix}
maps.

Secondly, we note the limitations of our simulations in approximating
both the shear and galaxy overdensity fields as Gaussian random fields
(GRFs) and that this is not a very good model, in particular for the
overdensity field. Defining the galaxy overdensity field as
\begin{equation}
\delta=\frac{N-\bar{N}}{\bar{N}}~,
\label{deltacounts}
\end{equation}
where $N$ is the number of galaxies in a map pixel and $\bar{N}$ is
the mean galaxy occupation number in a pixel, averaged across all map
pixels, we note that $\delta$ values constructed from a galaxy
position catalogue according to equation~(\ref{deltacounts}) must lie
in the range $-1 \leq \delta \leq \delta'_{\rm max}$, where
$\delta'_{\rm max}$ is some arbitrary maximum value. (The lower limit
is $-1$ since the minimum number of galaxies that can be contained
within a single pixel is $N = 0$). However, for a given set of theory model
power spectra ($C_\ell^{G_fG_f}, C_\ell^{G_sG_s}, C_\ell^{G_s\kappa_f}$, ), our GRF simulations produce
overdensity values in some other (symmetric) range $-\delta_{\rm max}
\lsim \delta \lsim \delta_{\rm max}$, where $\delta_{\rm max}$ is a
non-unity number which is determined by the amplitude of the model
power spectra.


To account for this discrpeancy and to properly connect our input
theoretical models with the galaxy-galaxy lensing statistics which we
measure from the simulated catalogues, we re-calibrate the simulated
overdensity fields by a factor $1/\delta_{\rm max}$ so that they lie
in the range $-1 \lsim \delta \lsim 1$. This allows us to populate the
map pixels in a self-consistent way so that the galaxy population of
every map pixel is strictly $\geq 0$ as required, subject to the
constraint that the total number of lens galaxies in the simulation is
kept fixed at the true value of $N^{\rm SDSS}_{\rm gal} = 38.5$ million. 
Our final theoretical prediction for the galaxy-galaxy lensing signal
as measured from the scaled simulation, and also taking into account
the pixelisation smoothing is then simply (c.f. equation~\ref{tangshearst})
\begin{equation}
\gamma_t(\theta)=\frac{1}{2\pi\delta_{\rm max}}\int _0 ^{\ell_{\rm max}} \ell \mathrm {d}\ell
F_{\ell} ^{2} C_{\ell}^{G\kappa} J_2({\ell \theta})~,
\label{pixtangshearst_mod}
\end{equation}
where $\ell_{\rm max}$ is the maximum multipole that was included in
the simulation.

\begin{figure*}
\subfigure{\includegraphics[width=5.5cm]{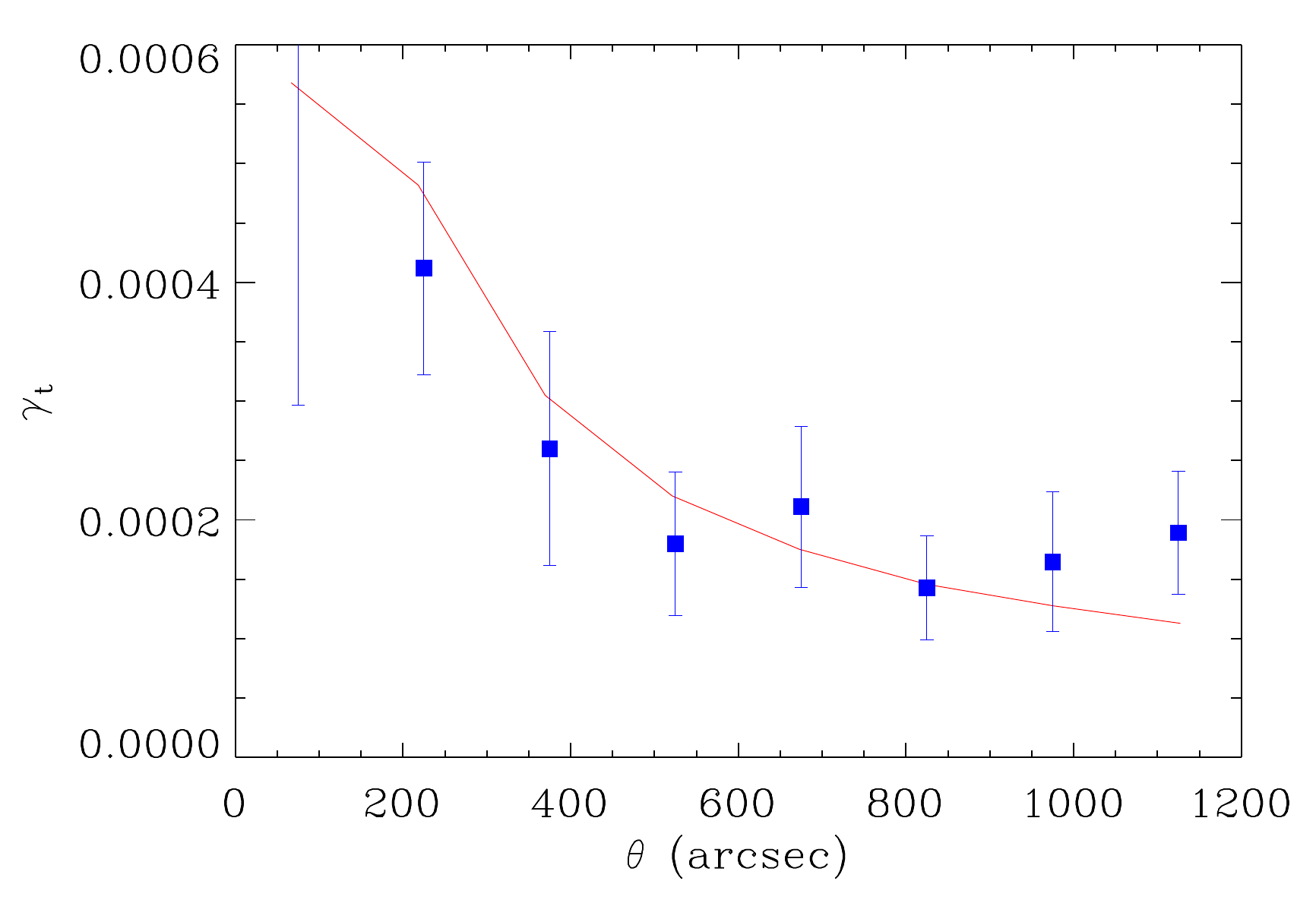}}\hspace{0.5em}
\subfigure{\includegraphics[width=5.5cm]{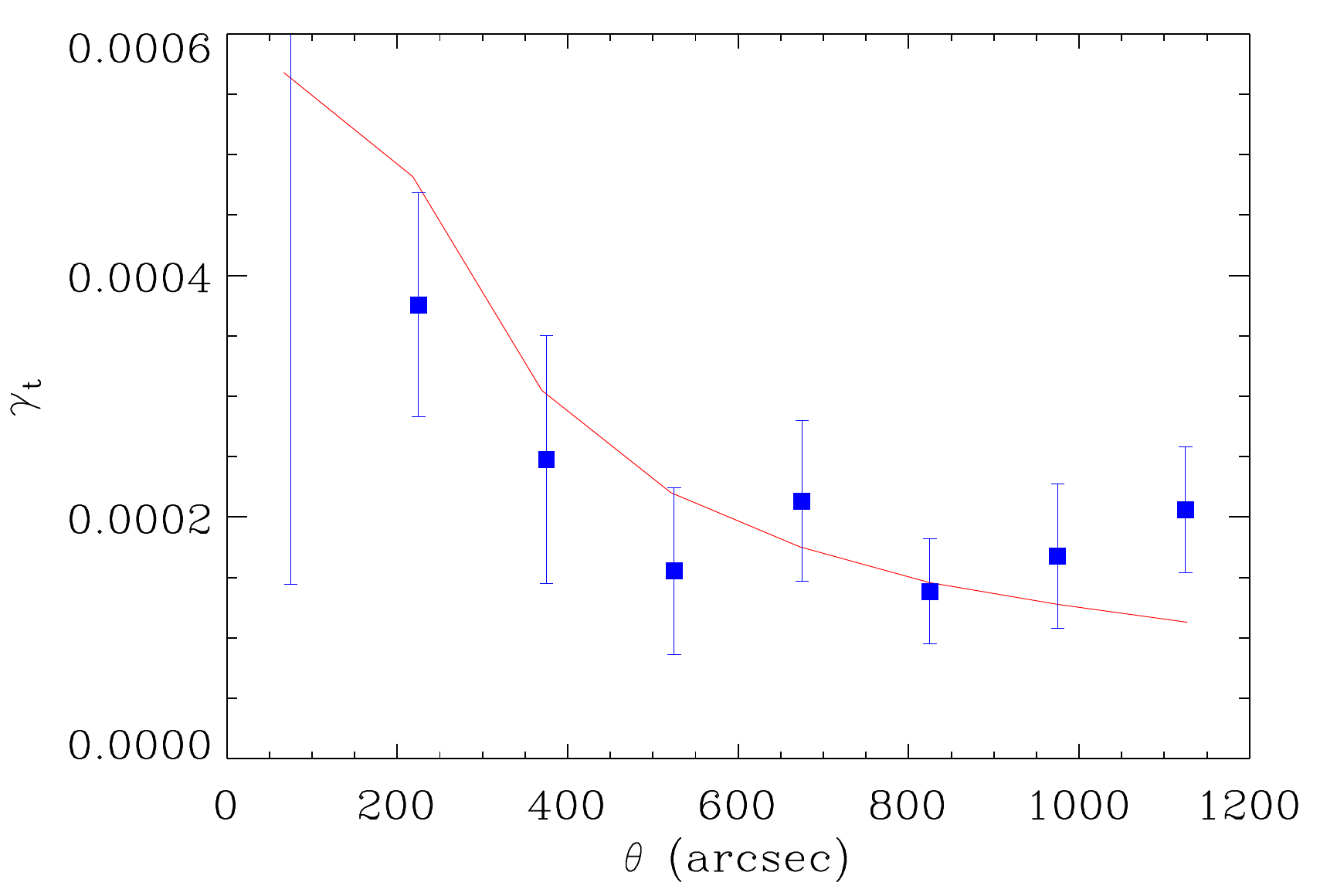}}\hspace{0.5em}
\subfigure{\includegraphics[width=5.5cm]{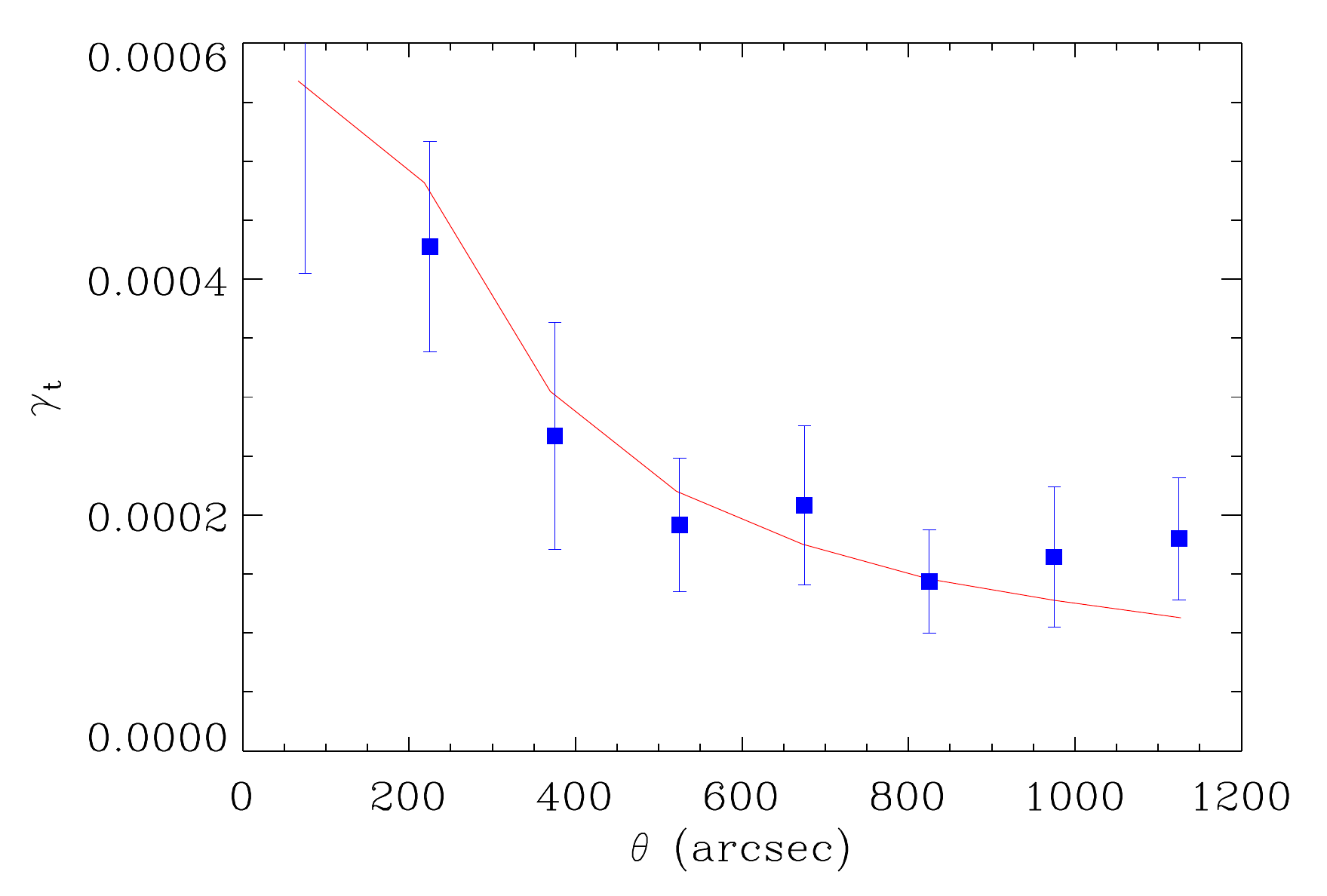}}

\caption[ The tangential shear signal measured from 100 simulations in
  the absence and presence of systematics and after the FIRST
  simulated shapes were corrected. Over-plotted (red line) is the
  input tangential signal to the simulations.]{From left to right are
  the mean recovered tangential shear signal measured from 100
  simulations in the absence (\emph{left panel}) and presence
  (\emph{centre panel}) of systematic effects, and after the FIRST
  simulated shapes were corrected (\emph{right panel}). Over-plotted
  as the red line in each case is the predicted tangential signal
  (equation~\ref{pixtangshearst_mod}).}
 \label{simulationsreswl}
\end{figure*}

Fig.\,\ref{simulationsreswl} shows the tangential weak lensing shear
signal recovered in the absence (left panel) and in the the presence
(centre panel) of the simulated FIRST systematics, and after shape
corrections were applied (right panel), averaged over the 100
Monte-Carlo (MC) simulations that we have performed. The three panels
show a very similar picture, although the points in the measurement
that was made using the uncorrected FIRST shapes seem to have a
slightly larger scatter around the predicted theory curve. However, in
all three cases, the measured tangential shear is consistent with the
input signal, while the rotated shear (not shown) was found to be
consistent with zero.

We note that the unbiased recovery that we observe in our simulations
is, in part, due to the fact that we do not include any position
correlations between our two simulated galaxy samples. This means that
the spurious position-shape correlations that were included in the
mock FIRST data will not propagate to the cross correlation
measurement between the simulated SDSS galaxy positions and the
simulated FIRST galaxy shapes. The cross-correlation tangential shear
signal in the presence of these small scale systematics in FIRST is
consequently unbiased and therefore the correction step does not
improve the outcome. 

Finally, we note that the uncertanties (which are calculated from the
run-to-run scatter amongst the 100 MC simulations) in all three cases
are also very similar. This is because at this point the results are
dominated by statistical errors due to the random intrinsic dispersion
in the FIRST galaxy shapes. Further investigation of this matter has
shown that by increasing the number of FIRST or SDSS sources (and
therefore decreasing the statistical uncertainties), the additional
scatter induced by the modelled contamination in the data is no longer
negligible (the error bars increase by a few percent).

\section{Real Data Measurements}\label{rdm}
In the previous section we have used simulations (which include a
model of the VLA beam systematics) to demonstrate that our analysis
pipeline is able to successfully recover a known galaxy-galaxy lensing
signal that was injected into the simulated FIRST galaxy shapes. We
now apply the same analysis pipeline to the real data. To estimate the
uncertainties in our measurements we once again use the scatter in the MC
simulations.

\begin{figure*}
\subfigure{\includegraphics[width=5.5cm]{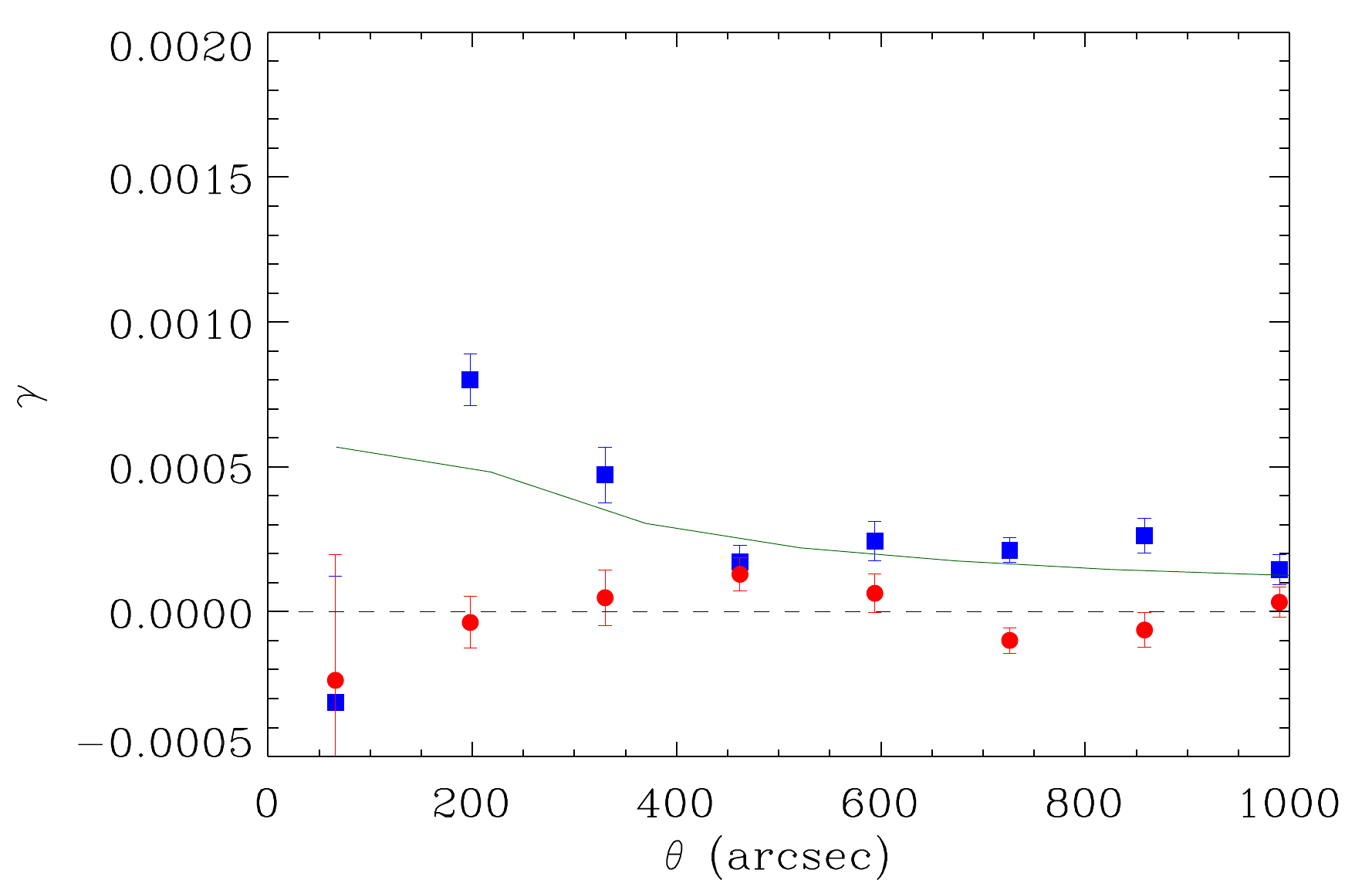}}\hspace{0.5em}
\subfigure{\includegraphics[width=5.5cm]{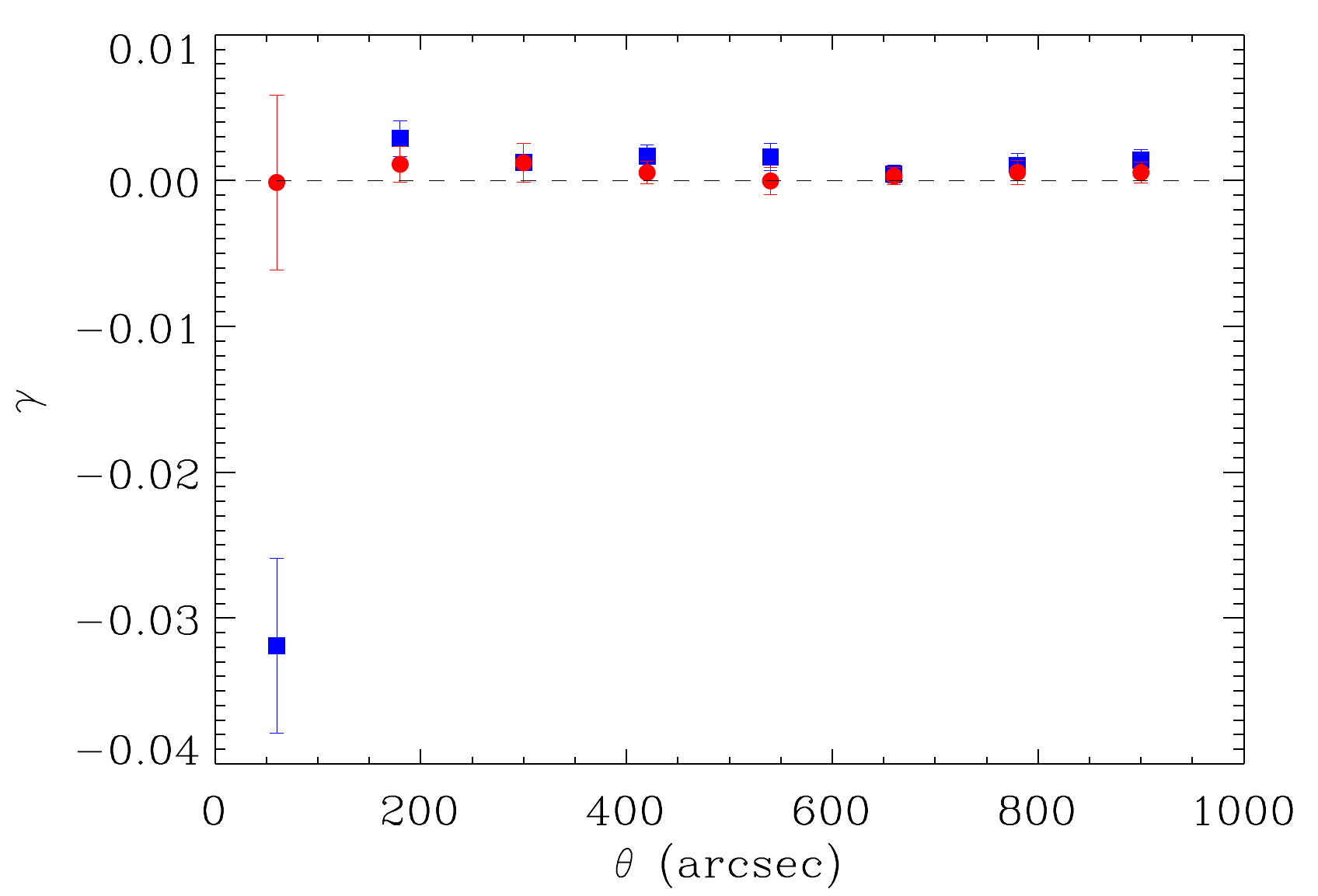}}\hspace{0.5em}
\subfigure{\includegraphics[width=5.5cm]{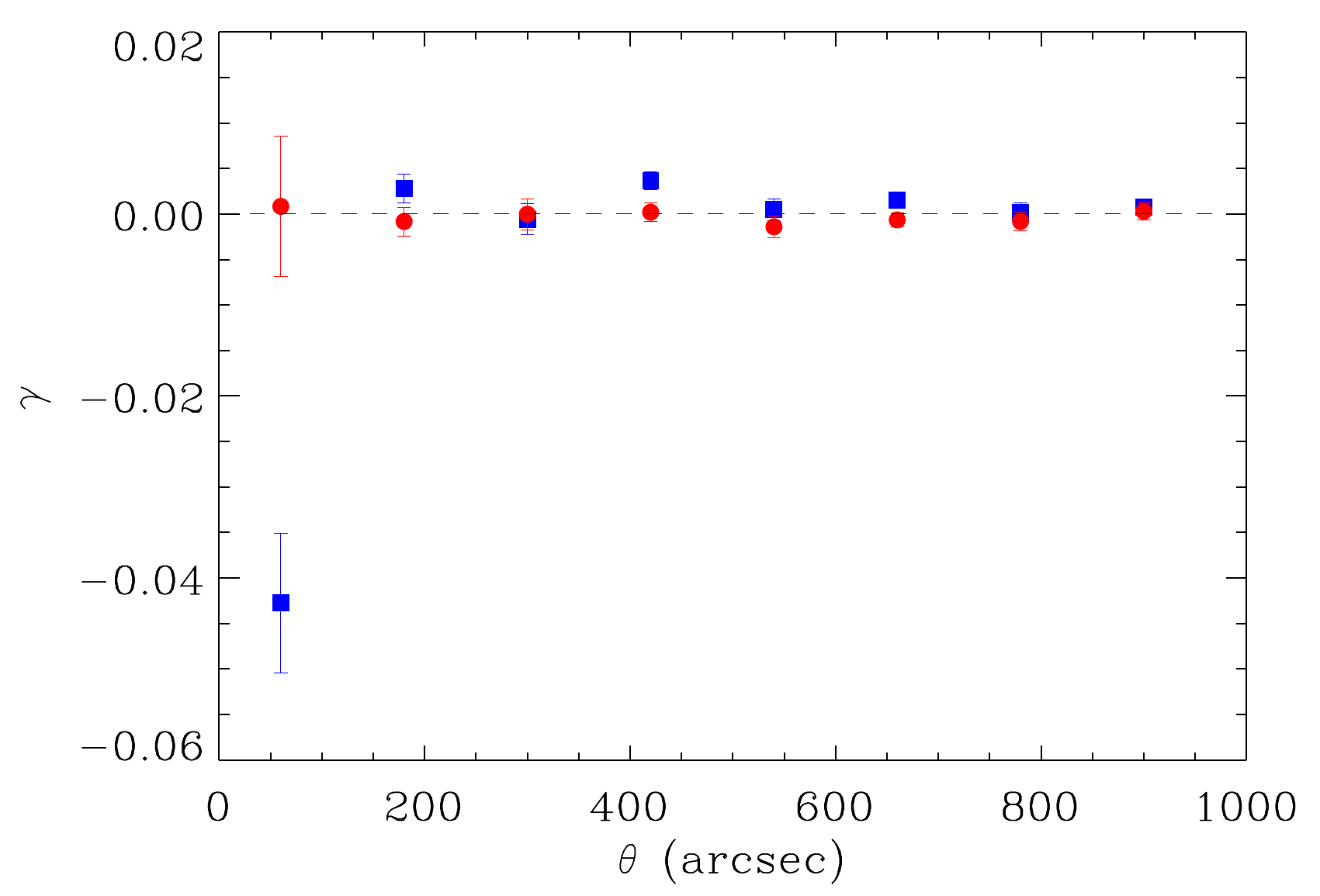}}
\subfigure{\includegraphics[width=5.5cm]{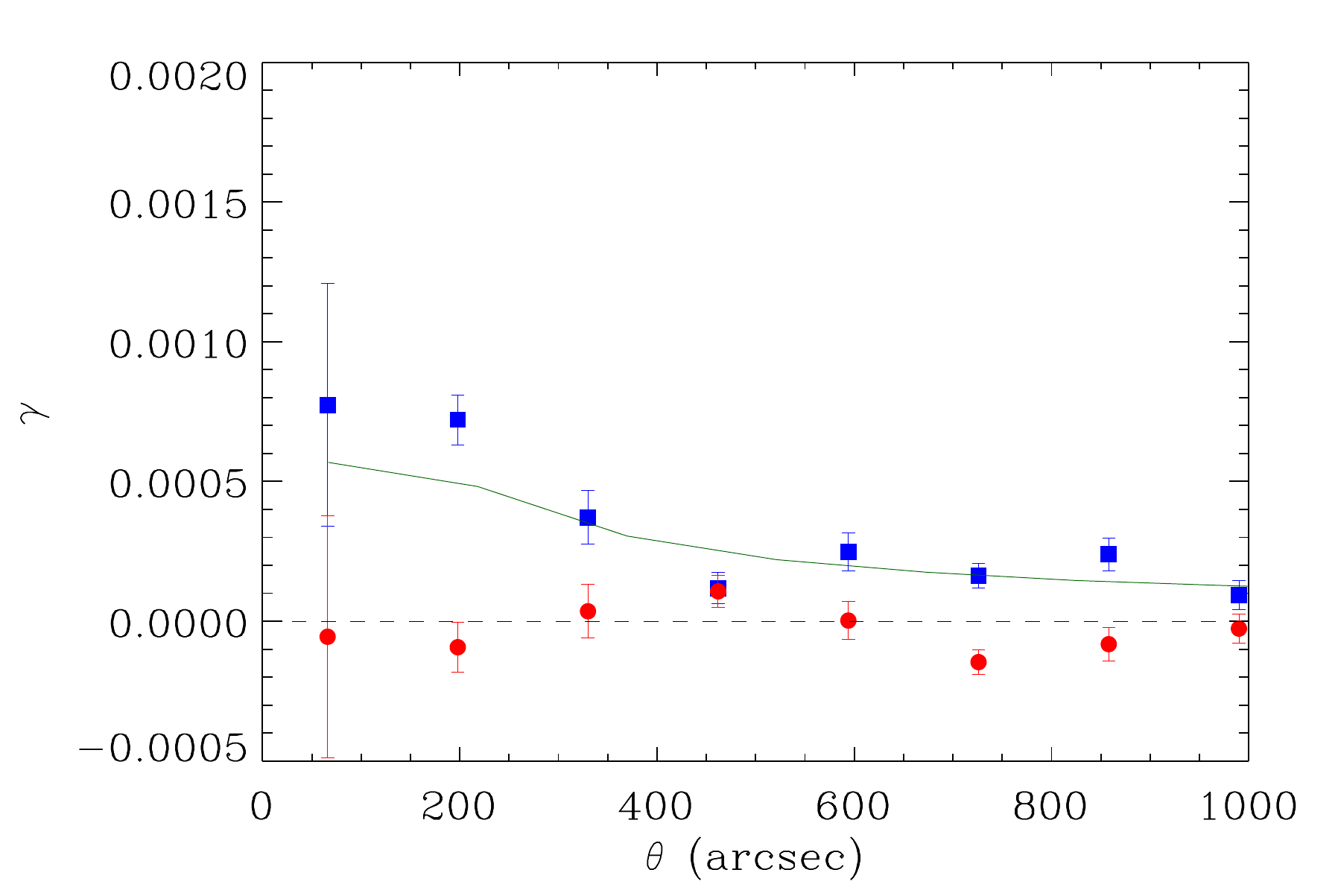}}\hspace{0.5em}
\subfigure{\includegraphics[width=5.5cm]{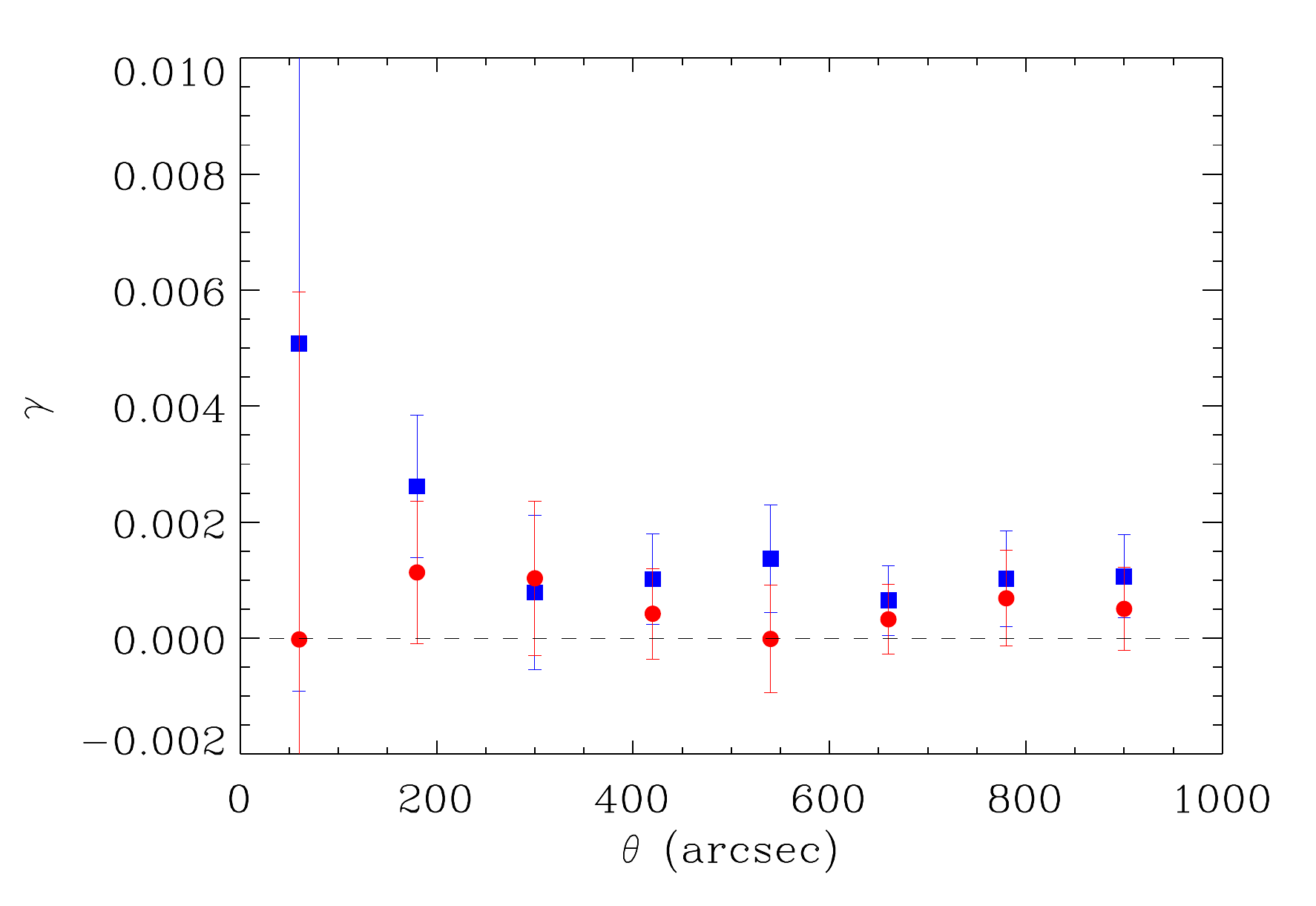}}\hspace{0.5em}
\subfigure{\includegraphics[width=5.5cm]{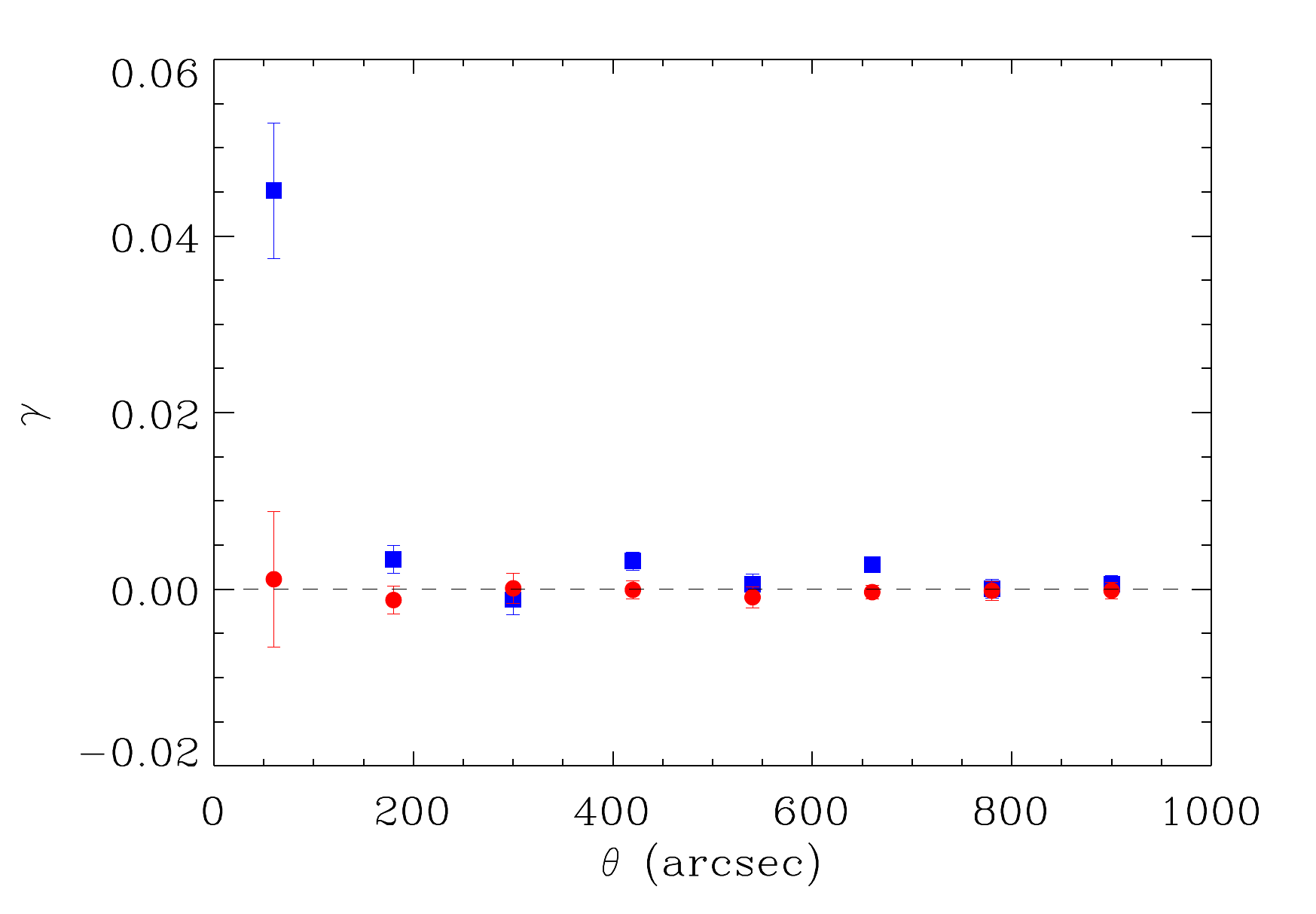}}
\caption{The measured tangential (blue squares) and rotated (red
  circles) shear using as lenses the SDSS complete catalogue
  (\emph{left panels}), the BCG sample (\emph{centre panels}) and the
  sample composed of the SDSS-FIRST matched objects (\emph{right
    panels}). The upper panels show measurements using the uncorrected
  FIRST galaxy shapes while the lower panels show the results after
  the shape corrections were applied. Over-plotted in the left panels
  is the theoretical tangential shear signal for a set of foreground
  and background sources lying at redshifts of
  $z^{1}_{\mathrm{median}}$=0.53 and $z^2_{\mathrm{median}}$=1.2
  respectively.}
 \label{realdatasig}
\end{figure*}

The galaxy-galaxy lensing signal measured from the real data (by
stacking the shapes of the FIRST galaxies around the positions of the
SDSS galaxies) is shown in the left hand panels of
Fig.~\ref{realdatasig}. When comparing the signals before and after
correcting the FIRST shapes (top and bottom panels respectively) one
can see that the FIRST beam systematic effect has corrupted the
tangential shear signal on scales $\theta \lesssim 200$
arcsec, while the rest of the data points (both for $\gamma_t$ and
$\gamma_r$) are almost unchanged. A similar picture emerges for the
cases where we consider the other two lens samples (i.e. the BCG and
SDSS-FIRST matched samples -- these are shown in the centre and right
hand panels of Fig.~\ref{realdatasig} respectively).  We note
that on application of our correction algorithm for the FIRST shape
systematics, the detected signal on scales $\theta < 200$
arcsec changes from a strong radial distortion signal (which is
consistent with being caused by the spurious systematics already
identified in the FIRST data) to a tangential distortion signal after
correction that is broadly consistent with theoretical expectations.

We find the tangential shear measurement made using the FIRST
corrected shapes for scales of $0 \leq \theta \leq 1200$ arcsec, to be
inconsistent with zero at the $\sim10\sigma$ level. To assess the
degree with which the measured signal is in agreement with the
cosmological model we calculate the $\chi ^2$ misfit statistic as
\be
\chi^2 = \sum_b (\hat{\gamma}_b^t - \gamma_b^{t,\rm th})^2 / \sigma^2_{\hat{\gamma}_b^t},
\label{eq:chi2}
\ee
where $\gamma^{t,\rm th}_b$ is the expected value of the tangential
shear in a given angular separation bin $b$ and
$\sigma_{\hat{\gamma}_b^t}$ is the errorbar calculated using the
scatter in the measurements of the equivalent angular separation bin
amongst the MC simulations.  For scales $\theta \gsim 200$ arcsec we
convert the misfit statistics to likelihood values for a model with
two degrees of freedom, the median redshift of the FIRST and SDSS
surveys.
We find that the measured signal is in agreement with the theoretical
predictions for a {\it Planck} cosmology and two galaxy populations with median
redshifts, $z^{\rm{SDSS}}_{\rm{m}}=0.53$ and
$z^{\rm{FIRST}}_{\rm{m}}=1.2$ at the 2$\sigma$ level. We omit the
measurements in the two lowest $\theta$-bins from the analysis as the
adopted Gaussian model for the galaxy-galaxy lensing signal is
unlikely to be valid on these small scales. Finally for the main SDSS
lens sample, we find that the rotated shear signal is consistent with
zero.

For the other two lens samples, we also
measure a tangential shear signal that is inconsistent with zero: at
$3.8\sigma$ for the BCG sample and at 9$\sigma$ for the FIRST-SDSS
matched galaxy lens sample (shown in the upper centre and upper right
panels of Fig.\,\ref{realdatasig} respectively). The rotated shear
signal for the SDSS-FIRST matched galaxy lens sample is consistent with zero. The
measured rotated shear for the BCG lens sample, although lower than the
measured tangential shear, is inconsistent with zero. As expected, the
tangential shear measured around the BCG lens sample is $\sim$1 order
of magnitude larger than that measured
using the main SDSS DR10 catalogue. Perhaps unexpectedly though, the shear signal
detected around the SDSS-FIRST matched objects is even higher than that
measured around the BCGs. It also appears to have the steepest slope
out of the three cases as it becomes consistent with zero for scales
$\theta \gtrsim 150$ arcsec. This scale, assuming a median redshift for the
lens sample of $z^{\rm matched}_{\rm m}$=0.57, corresponds to
  a radius of $R \simeq 1$ Mpc.

\subsection{Residual Systematics Test Measurements}
To further assess the validity of the results we perform a set of
measurements on the real data which are designed to reveal residual
systematics that may remain in the data after application of our
correction algorithm. For all of these measurements we assign
errorbars by processing the simulated datasets in exactly the same way
as the real data and then calculating the scatter in the measurements
from the MC simulations.

We have performed the following null tests for each of the three lens
samples that we have considered:

\begin{enumerate} 
\item \textit{\textbf{North-South}}: we split the lens sample and the
  FIRST data into roughly equal North and South samples. We then
  measure the galaxy-galaxy lensing signal from each sample
  separately. Finally, we subtract the signal between the two
  measurements.
\item \textit{\textbf{East-West}}: Same as test (i) but splitting the
  data into a West and an East component.
\item \textit{\textbf{Random lens}}: Randomly splitting the lens
  samples into two equal subsets. For each subset the galaxy-galaxy
  lensing signal is measured. Finally we subtract the two
  measurements.
\item \textit{\textbf{Random lens position}}: We randomly select
  positions on the sky to match the number of lenses in each lens
  group. We then use those positions as central stacking points and
  measure the mean stacked galaxy-galaxy lensing signal around those
  points.
\end{enumerate} 

\begin{figure*}
\subfigure{\includegraphics[width=5.5cm]{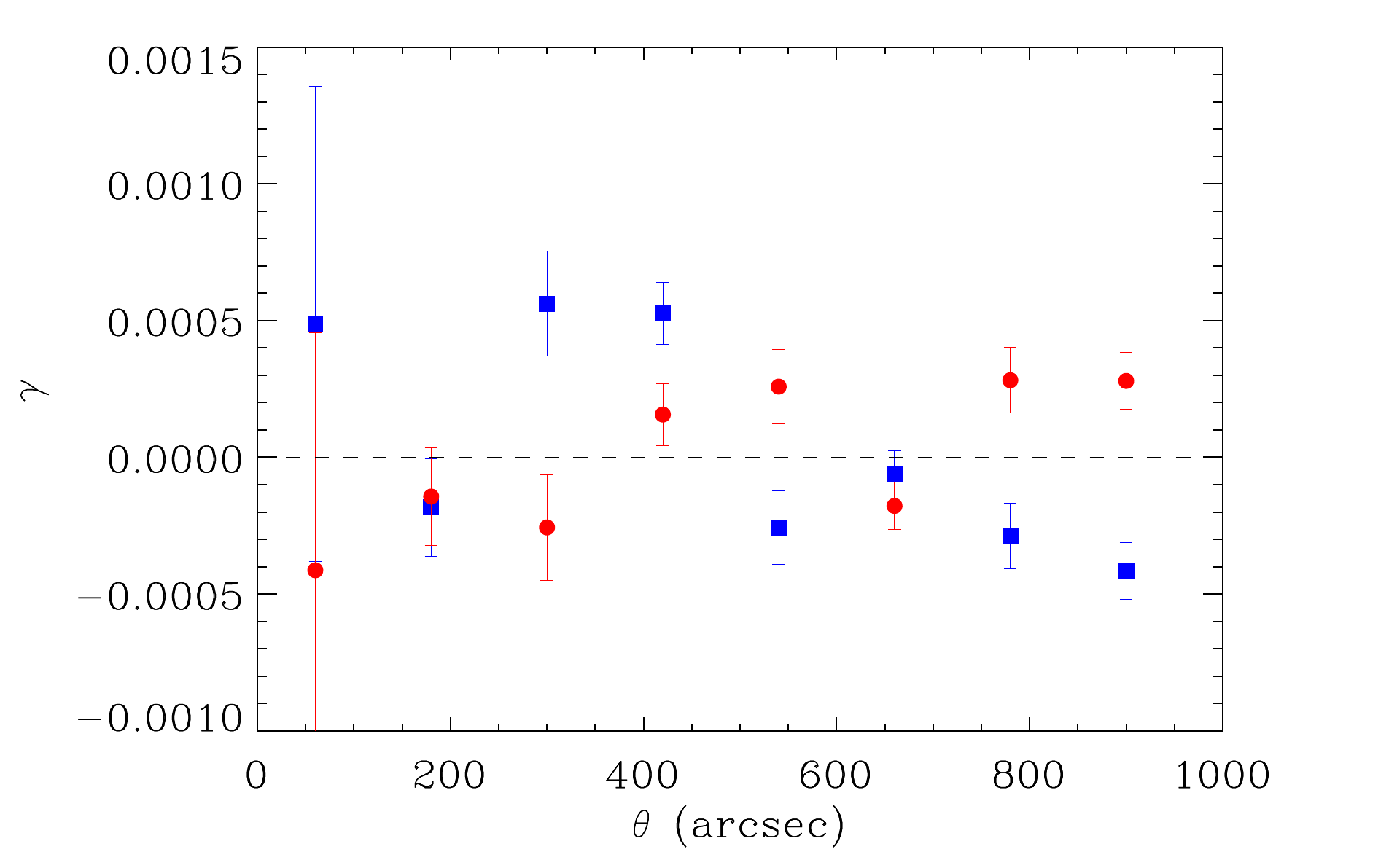}}\hspace{0.5em}
\subfigure{\includegraphics[width=5.5cm]{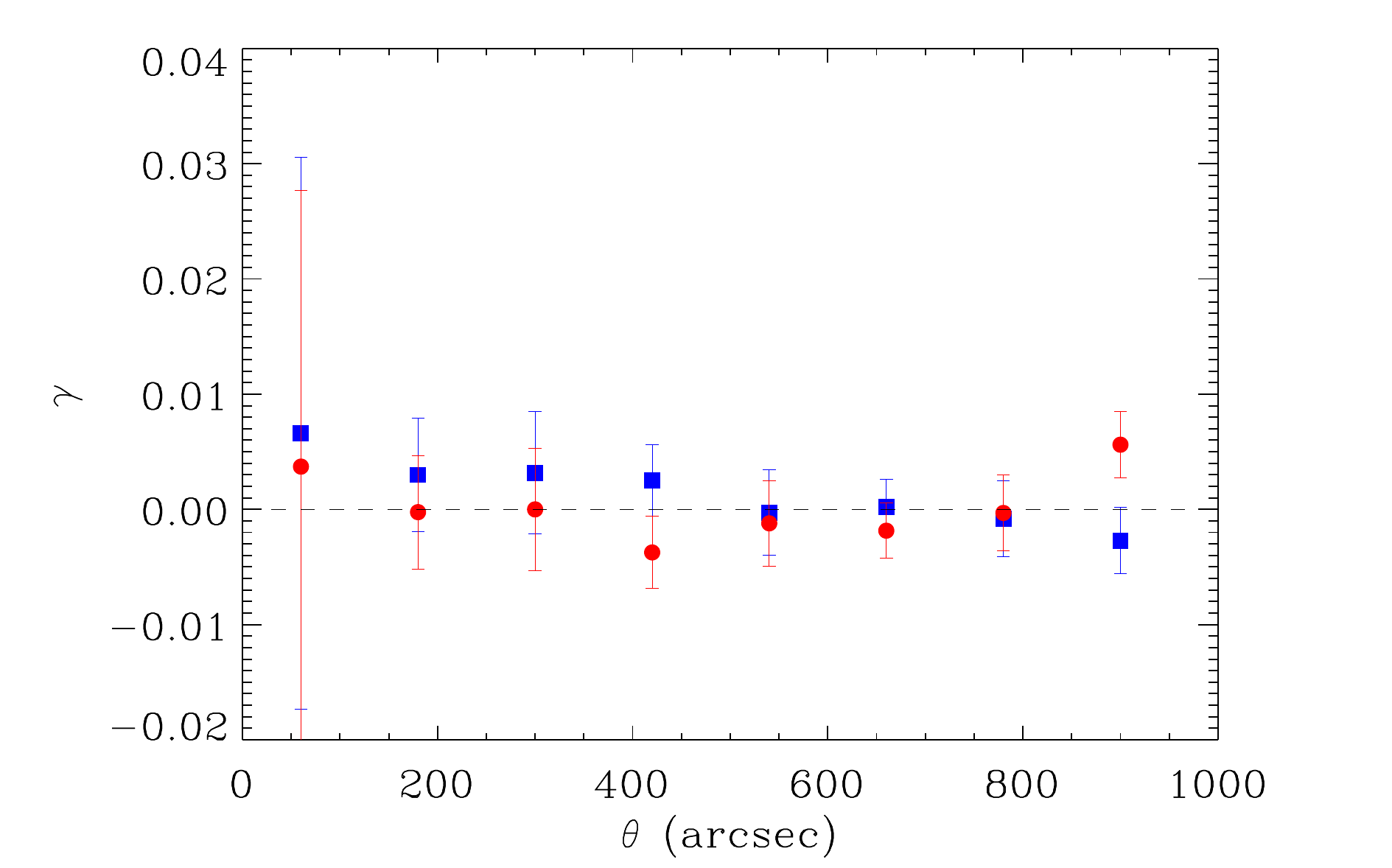}}\hspace{0.5em}
\subfigure{\includegraphics[width=5.5cm]{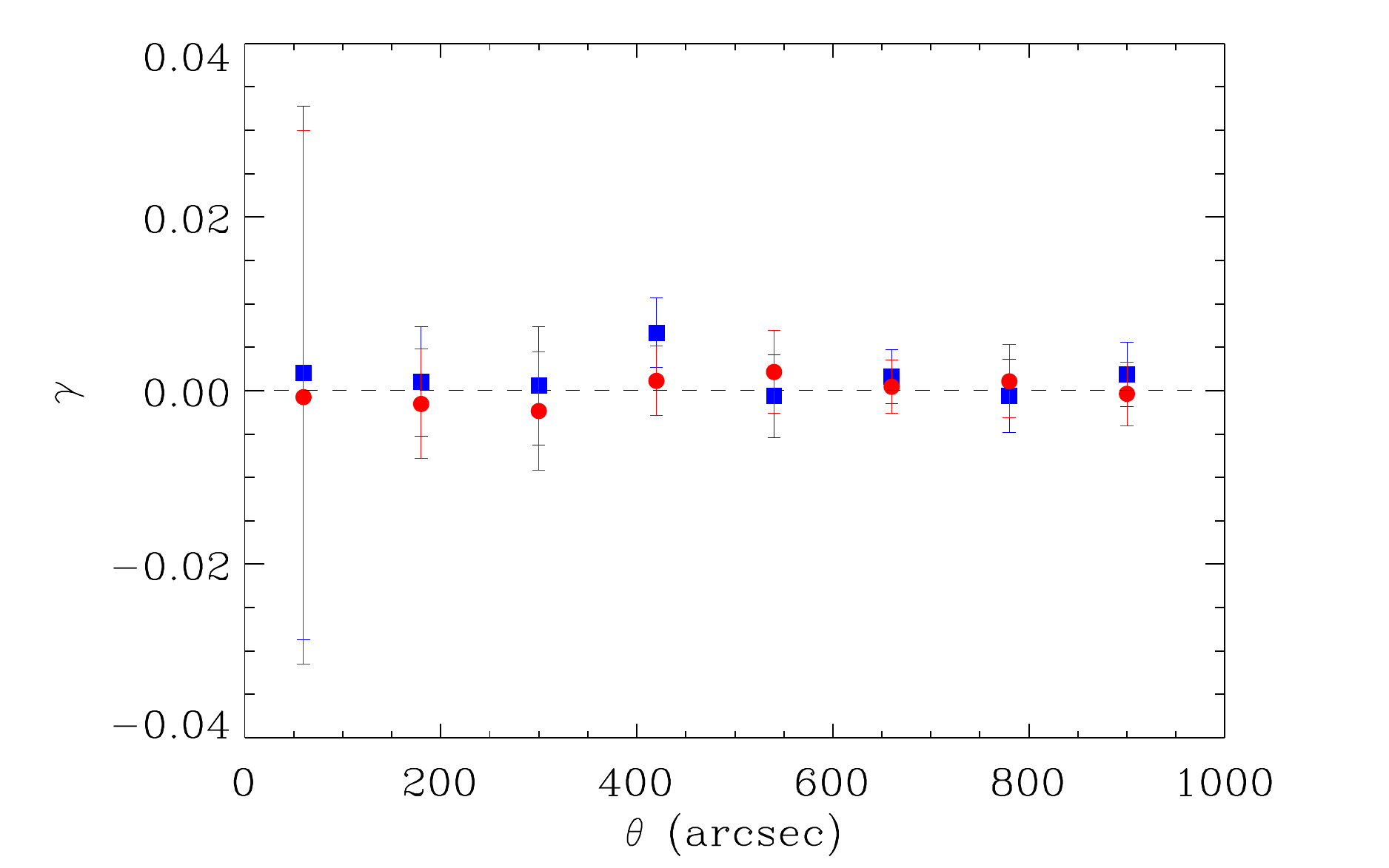}}\hspace{0.5em}
\subfigure{\includegraphics[width=5.5cm]{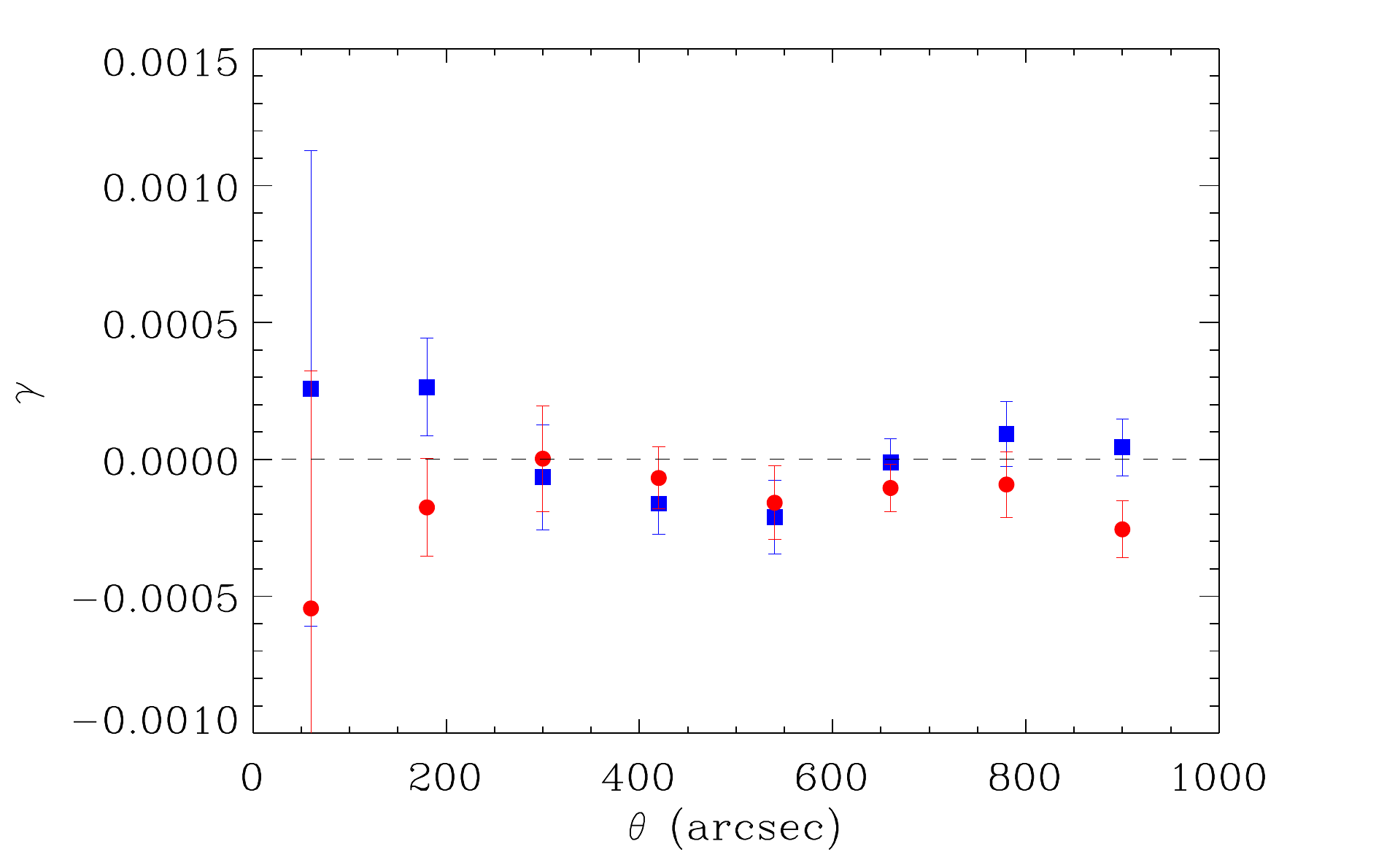}}\hspace{0.5em}
\subfigure{\includegraphics[width=5.5cm]{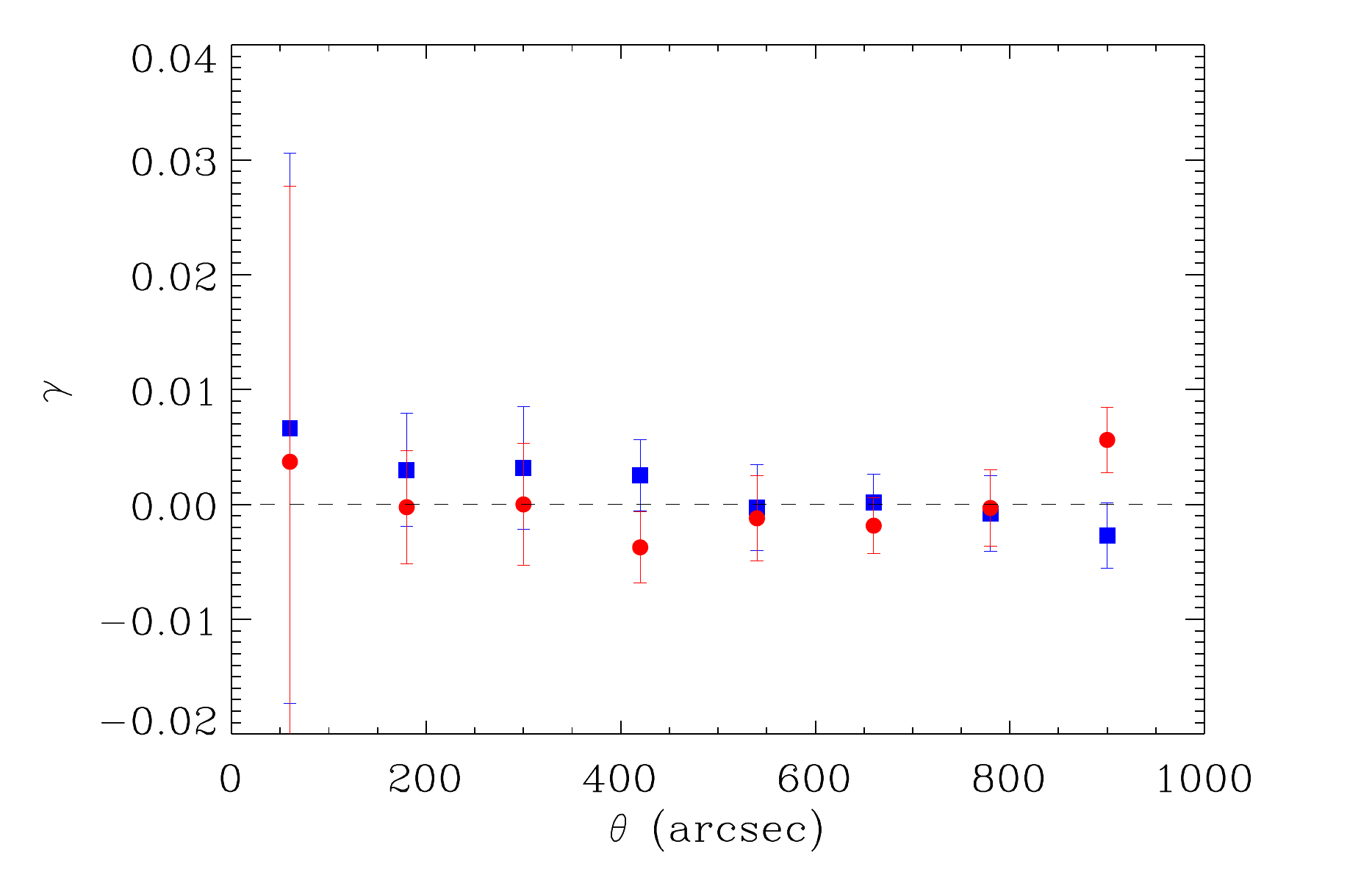}}\hspace{0.5em}
\subfigure{\includegraphics[width=5.5cm]{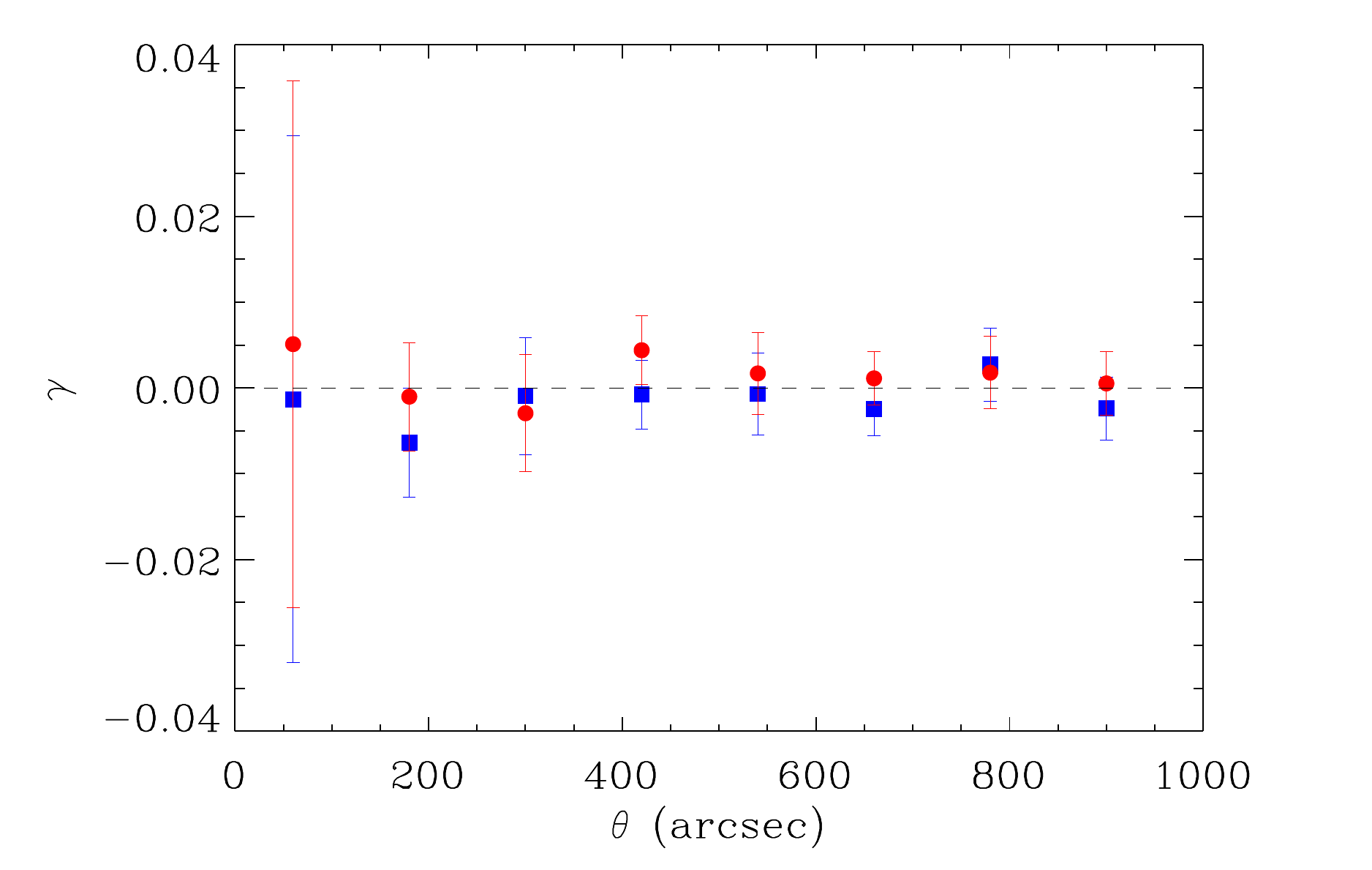}}\hspace{0.5em}
\subfigure{\includegraphics[width=5.5cm]{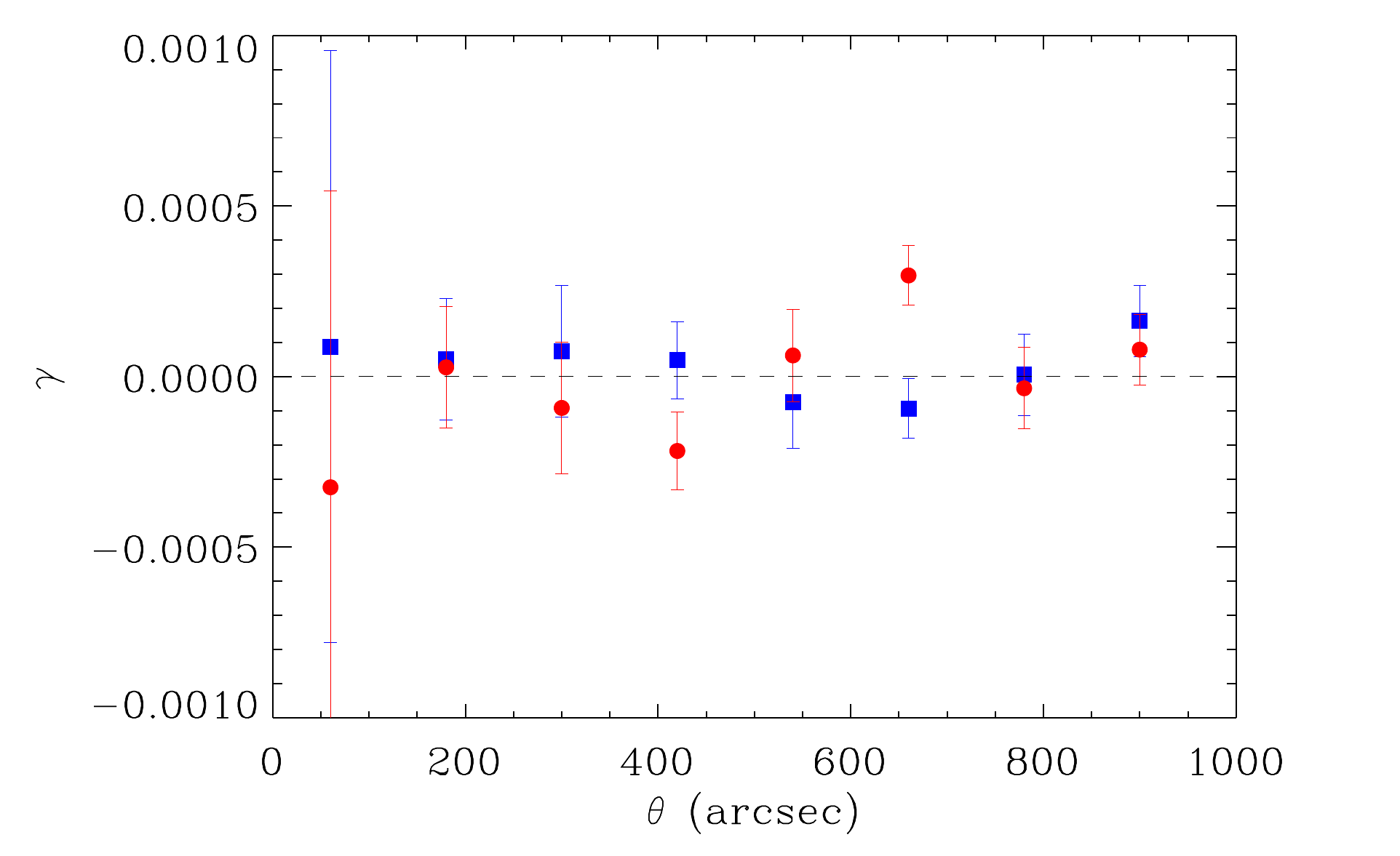}}\hspace{0.5em}
\subfigure{\includegraphics[width=5.5cm]{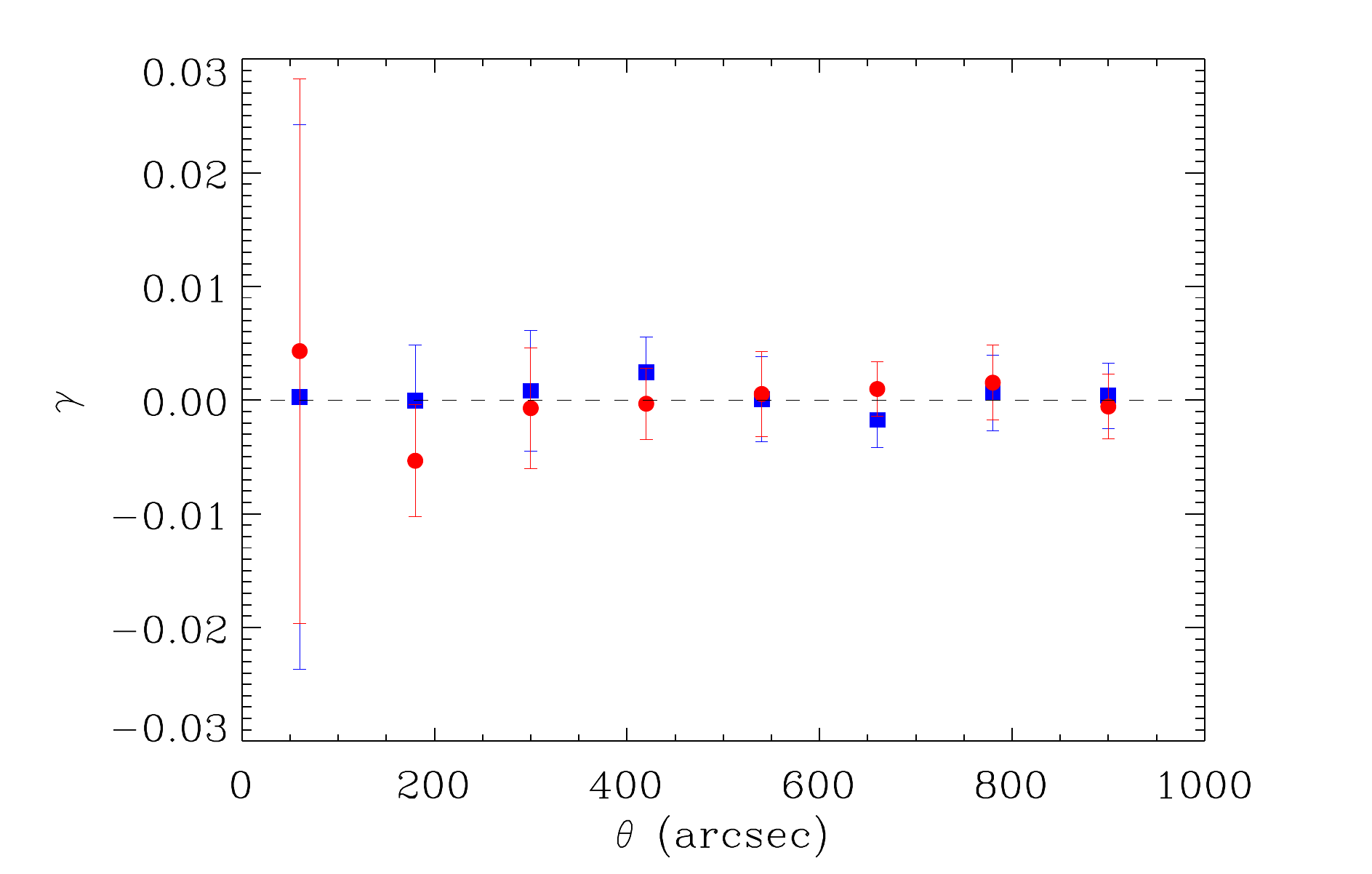}}\hspace{0.5em}
\subfigure{\includegraphics[width=5.5cm]{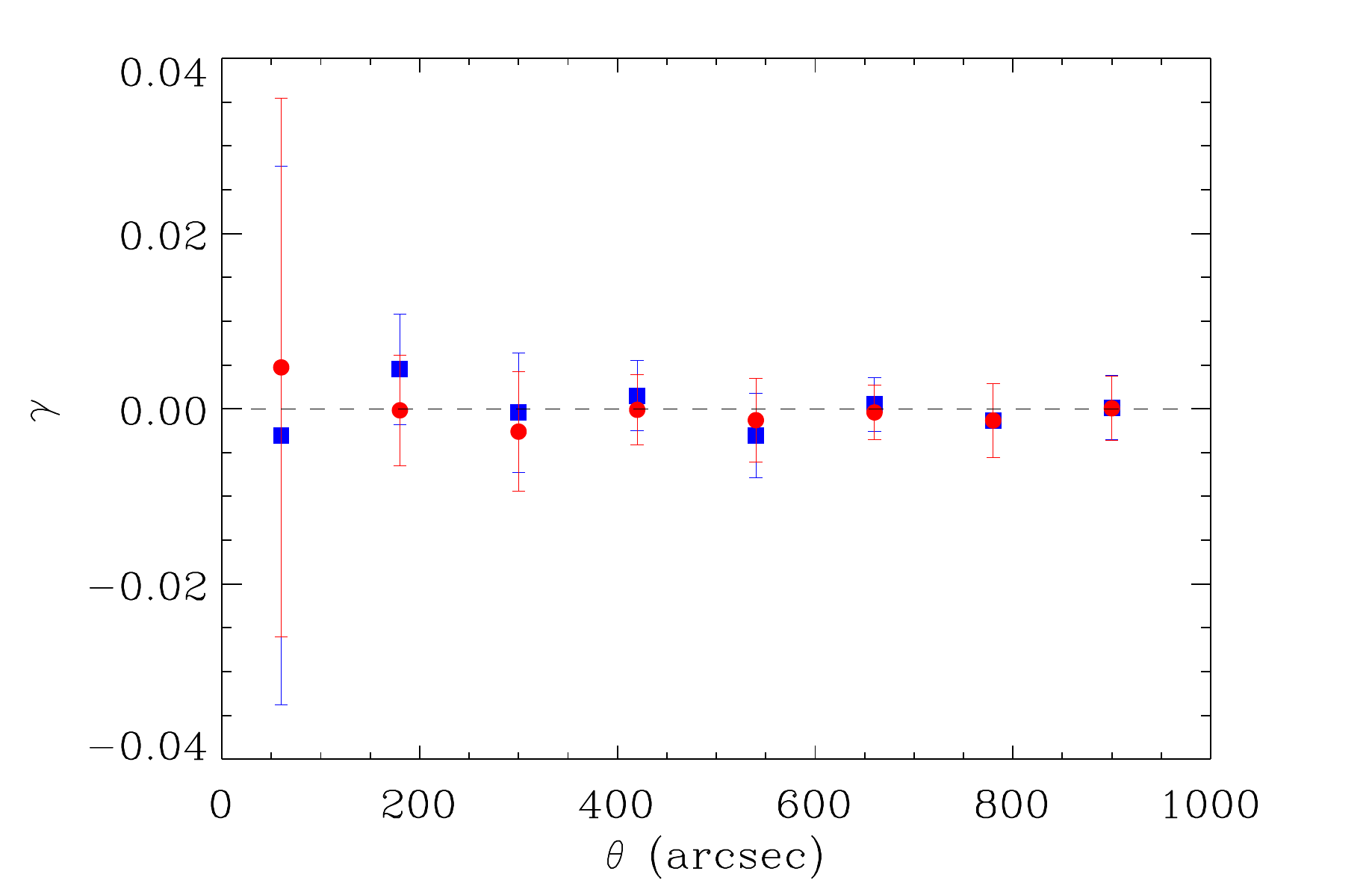}}\hspace{0.5em}
\subfigure{\includegraphics[width=5.5cm]{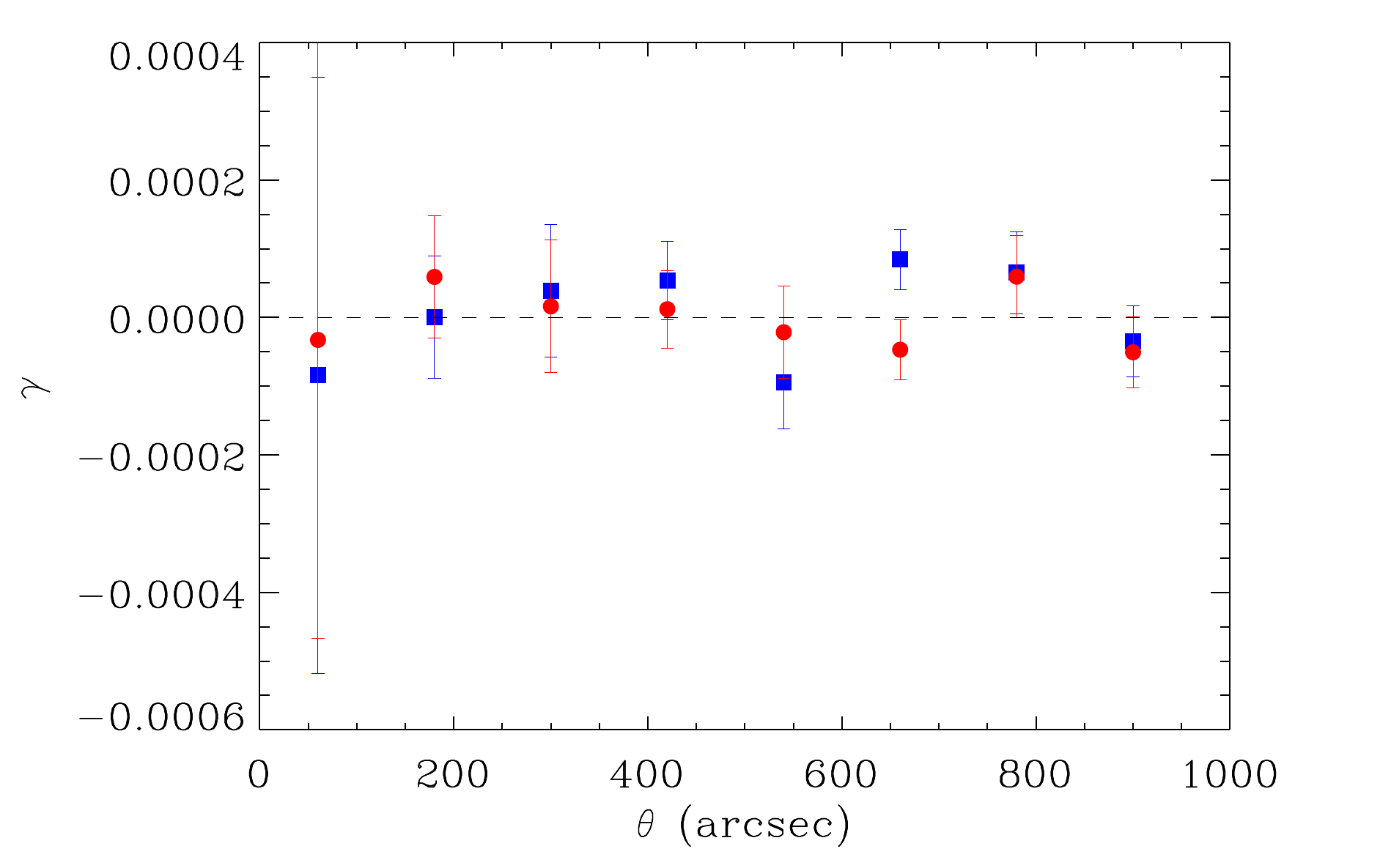}}\hspace{0.5em}
\subfigure{\includegraphics[width=5.5cm]{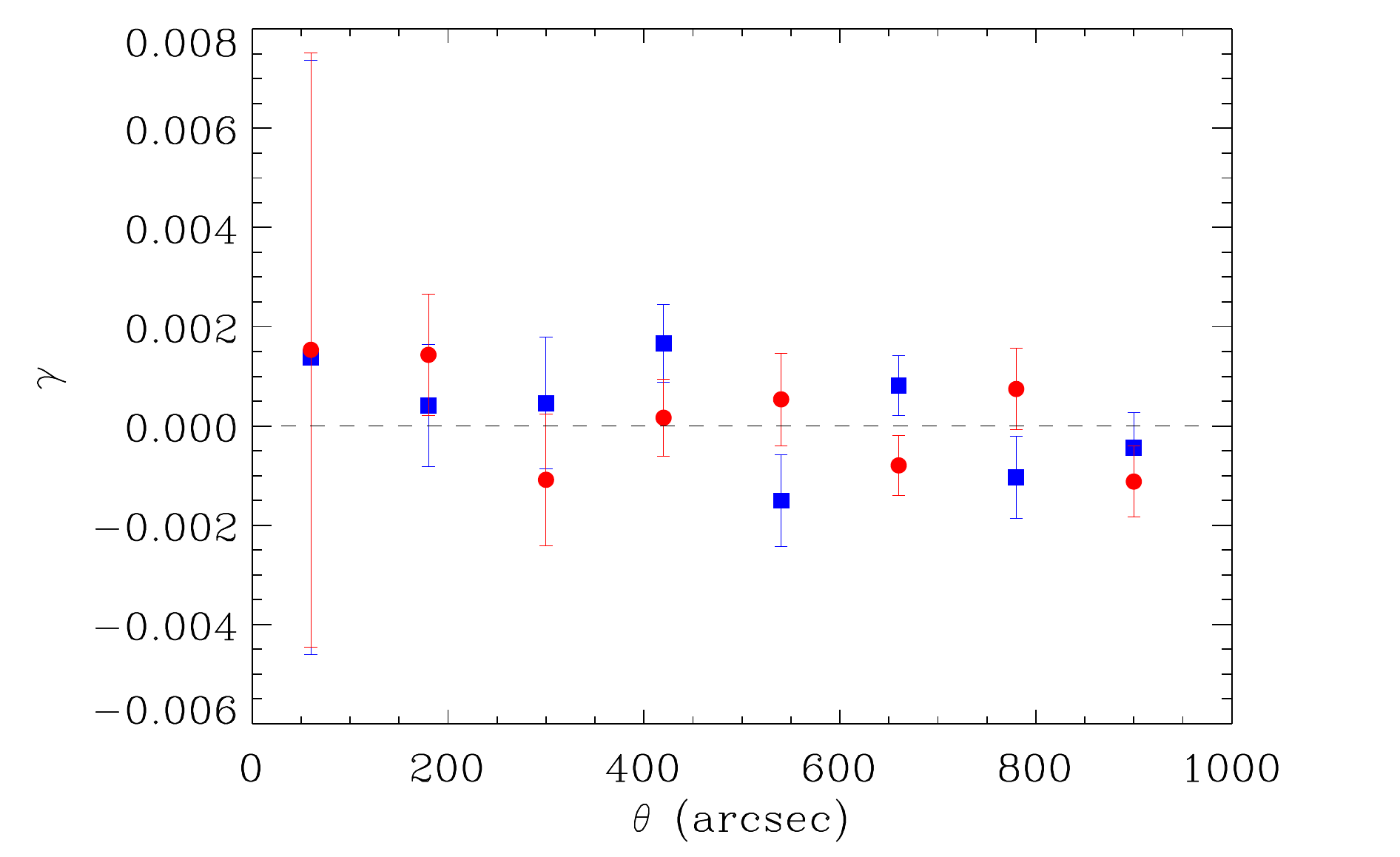}}\hspace{0.5em}
\subfigure{\includegraphics[width=5.5cm]{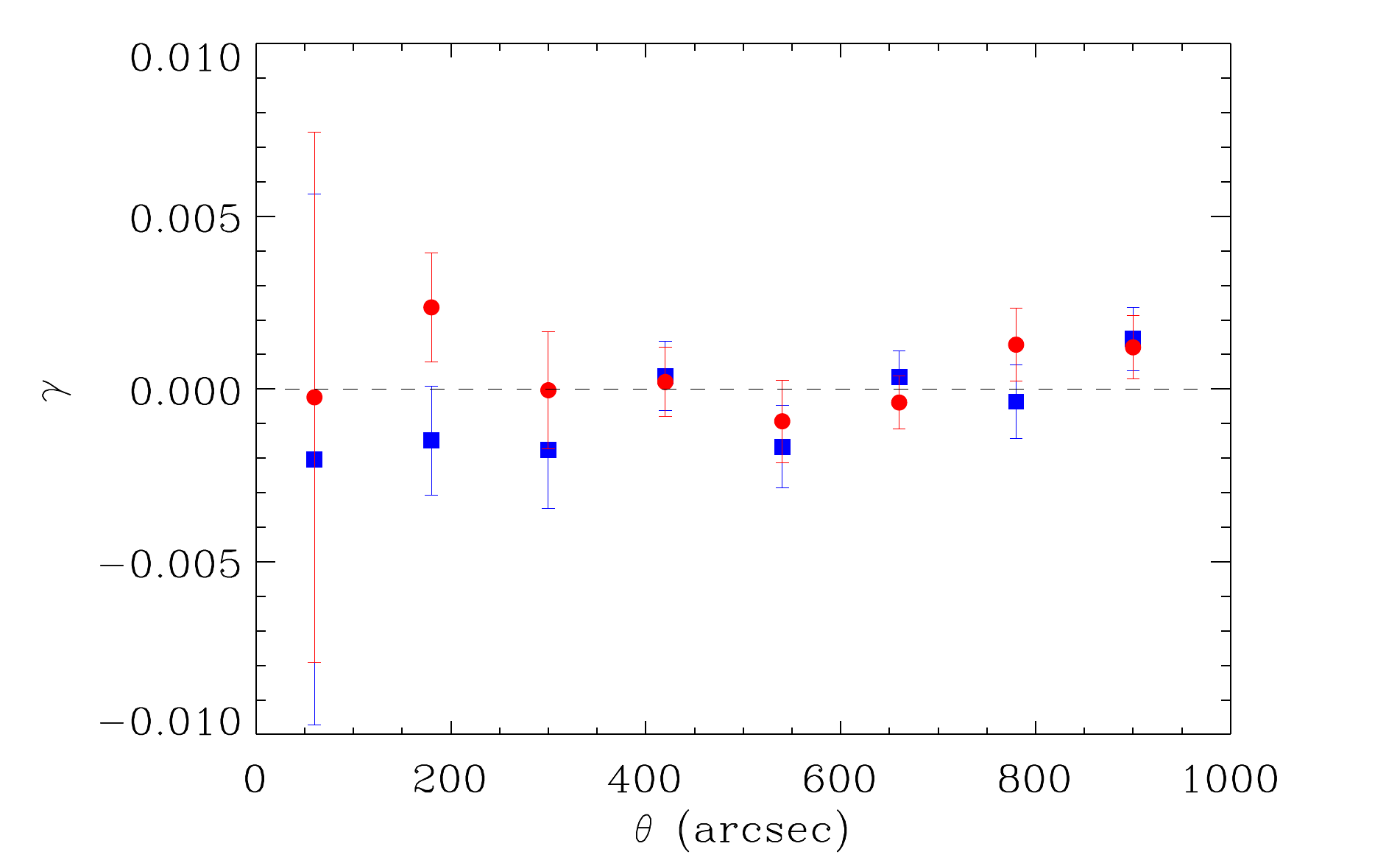}}
 \caption{Form left to right are the null tests conducted using the
   full SDSS sample (\emph{left panels}), the BCG lens sample
   (\emph{centre panels}) and the lens sample composed of the
   SDSS-FIRST matched objects (\emph{right panels}). From top to
   bottom the panels show the \emph{North-South}, \emph{West-East}, \emph{Random lens} and
   \emph{Random lens position} tests respectively. Blue squares and red
   circles show the measured tangential and rotated shear
   respectively.}
 \label{nulltestswl}
\end{figure*}

The results of these tests are shown in Fig.\,\ref{nulltestswl}. In
all cases the results, both in the tangential and the rotated
direction, are consistent with zero. The scatter of the points both
for the tangential and the rotated shear signal for the
\textit{\textbf{North-South}} test conducted on the SDSS full sample
seems to be larger than for the rest of the tests, but still no
obvious coherent signal could be detected. The tests therefore reveal
no major issues with the analysis that was followed.

\begin{figure} 
\centering
\includegraphics[width=8.5cm]{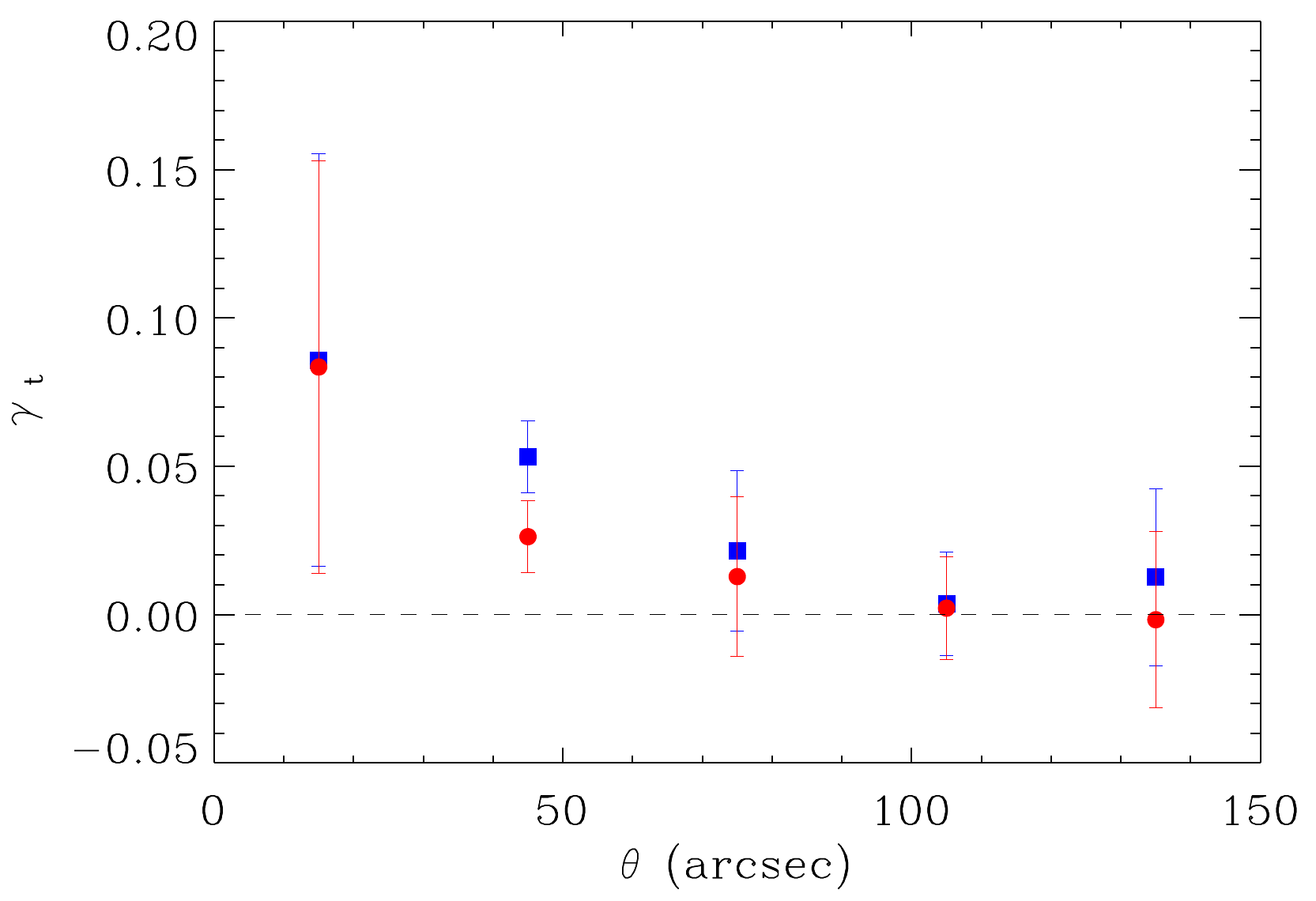}
\caption{The tangential shear signal measured using the FIRST selected
  sources as background objects and the SDSS-FIRST matched objects
  with $z_{low} <  1$ (blue squares) and
  $z_{ high} > 1$ (red circles) as lenses.}
\label{syst_SDSSdr10allFIRST_FIRST005_zdep}
\end{figure}

In addition to these null tests we have also looked into the redshift
dependence of the signal that was measured using the SDSS-FIRST
matched objects and the FIRST shapes. In this test we look on angular
scales where there is a significant non-zero tangential shear
signal ($\theta < 150$ arcsec). We compare the measured tangential shear
signal around the SDSS-FIRST sources that have redshifts
$z_{low} < 1$ and $z_{high} > 1$. The results (see
Fig.\,\ref{syst_SDSSdr10allFIRST_FIRST005_zdep}) although noisy, suggest
that the signal decreases when one changes lens samples from the low-redshift
sub-sample to the high-redshift sub-sample. This is consistent with the expected
behaviour for a real shear signal for which the strength of the
signal is expected to decrease as a function of the lens redshifts. 

\begin{figure*}
\subfigure{\includegraphics[width=8.5cm]{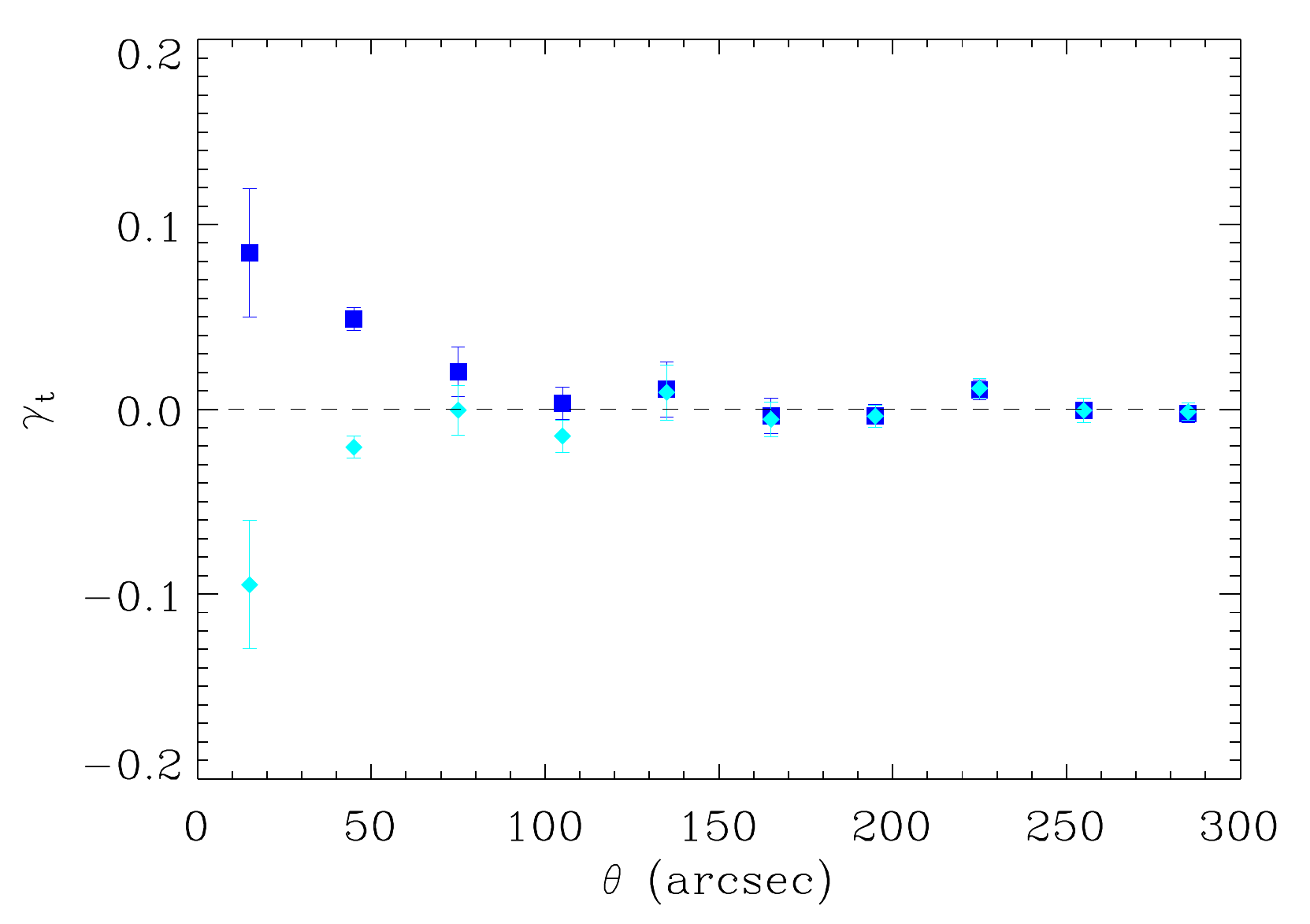}}\hspace{0.5em}
\subfigure{\includegraphics[width=8.5cm]{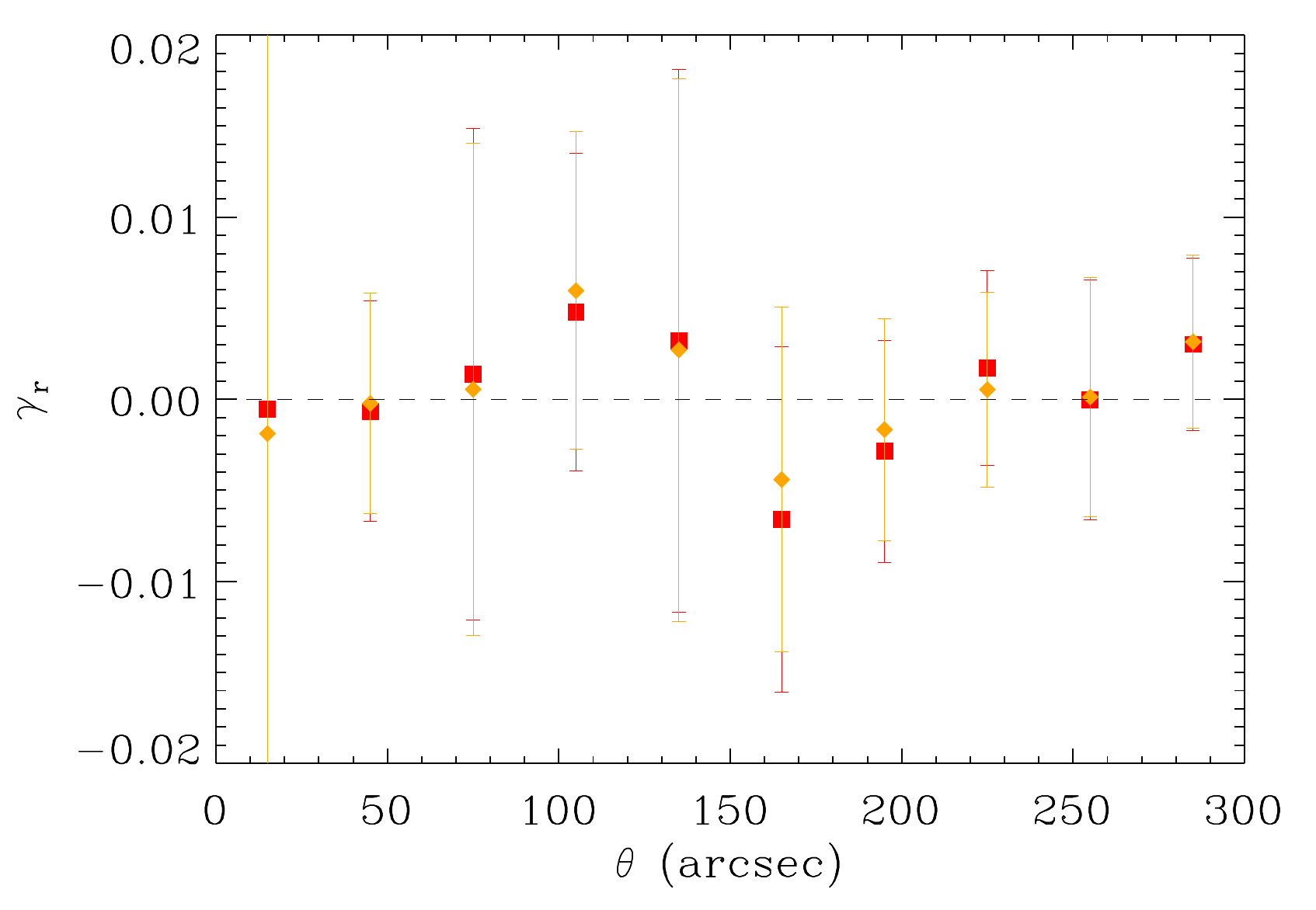}}
\caption{Galaxy-galaxy lensing measurements using the
  SDSS-FIRST matched objects as lenses and the selected FIRST sources
  as background objects. The tangential shear prior to (cyan circles) and
  after shape corrections (blue squares) is shown in the left panel
  and the rotated shear prior to (orange circles) and after shape
  corrections (red squares) is shown in the right panel.}
\label{fig:SDSSdr10FIRSTgtr}
\end{figure*}

Finally, we look at the signal dependence as a function of angular
scale $\theta$ before and after shape corrections for when the
SDSS-FIRST matched objects are used as lenses. We primarily focus on
this sample because a galaxy-galaxy lensing measurement, made by
stacking the FIRST shapes around the positions of this group, will
more likely be biased if the shape correction algorithm was not
successful (since all of the lenses in this sample are bright in the
radio). The results of this invesitgation are shown in
Fig.~\ref{fig:SDSSdr10FIRSTgtr}. As expected, the tangential shear
signal for this lens group prior to correcting the shapes of the FIRST
sources is negative for $\theta \leq 200$ arcsec. This signal,
although negative in absolute value, is smaller in amplitude than the
spurious signal detected when the complete FIRST catalogue was used as
central objects (see left panel of
Fig.\,\ref{syst_FIRSTrandompick_FIRST005sidelobe_alltog}). This
suggests that a strong positive cosmological tangential shear is also
present competing with the spurious negative one. The measured
tangential component of the shear after shape corrections are applied
is positive but has a steep slope as it becomes consistent with zero
for $\theta \gsim 150$ arcsec. 
The measured rotated shear signals for the two cases are consistent
with each other and they are also consistent with zero (see right
panel of Fig.\,\ref{fig:SDSSdr10FIRSTgtr}). All of these tests
show no obvious evidence of residual systematics in the data after the
shape correction step was performed.

\subsection{Constraints on the Properties of the Lensing Sources}
Having demonstrated the credibility of our results we now fit the two
dark matter halo models described in Section~\ref{dmhm} to the measured tangential shear to constrain
the ensemble mass $M_{200}$, the radius $R_{200}$, the velocity
dispersion $\sigma_u$ and the Einstein radius $\theta_{\mathrm{E}}$
for each of the lens samples that we have investigated. To fit the data
both models should be provided with the median redshifts of the lens
and background source populations. Additionally, for the NFW model,
the lens concentration factor can either be supplied or left to be
constrained by the data. Initially we choose to adopt the value for
this parameter drawn from \citet{bullock2001}. Subsequently we allow
the concentration factor to vary and we compare the returned $\chi^2$
and parameter values. The values for the sources' redshifts and
concentration factors that were extracted from the literature to fit
the three detected shear signals are summarised in
Table\,\ref{fittingpar}.

\begin{table}
\caption{The lens and background objects redshifts and concentration
  factor values used to fit the data.}
\begin{center}
\begin{tabular}{c|c|c|c}
\hline
 & SDSS & BCGs & SDSS-FIRST \\
 \hline
$z_{\rm{lens}}$ & 0.53 & 0.37 & 0.57 \\
\hline
$z_{\rm{BG sources}}$ & 1.2 & 1.2 & 1.2 \\
\hline
$c_f$ & 10 & 7 & 7  \\
\hline
\end{tabular}
\end{center}
\label{fittingpar}
\end{table}

\begin{figure*}
\subfigure{\includegraphics[width=5.5cm]{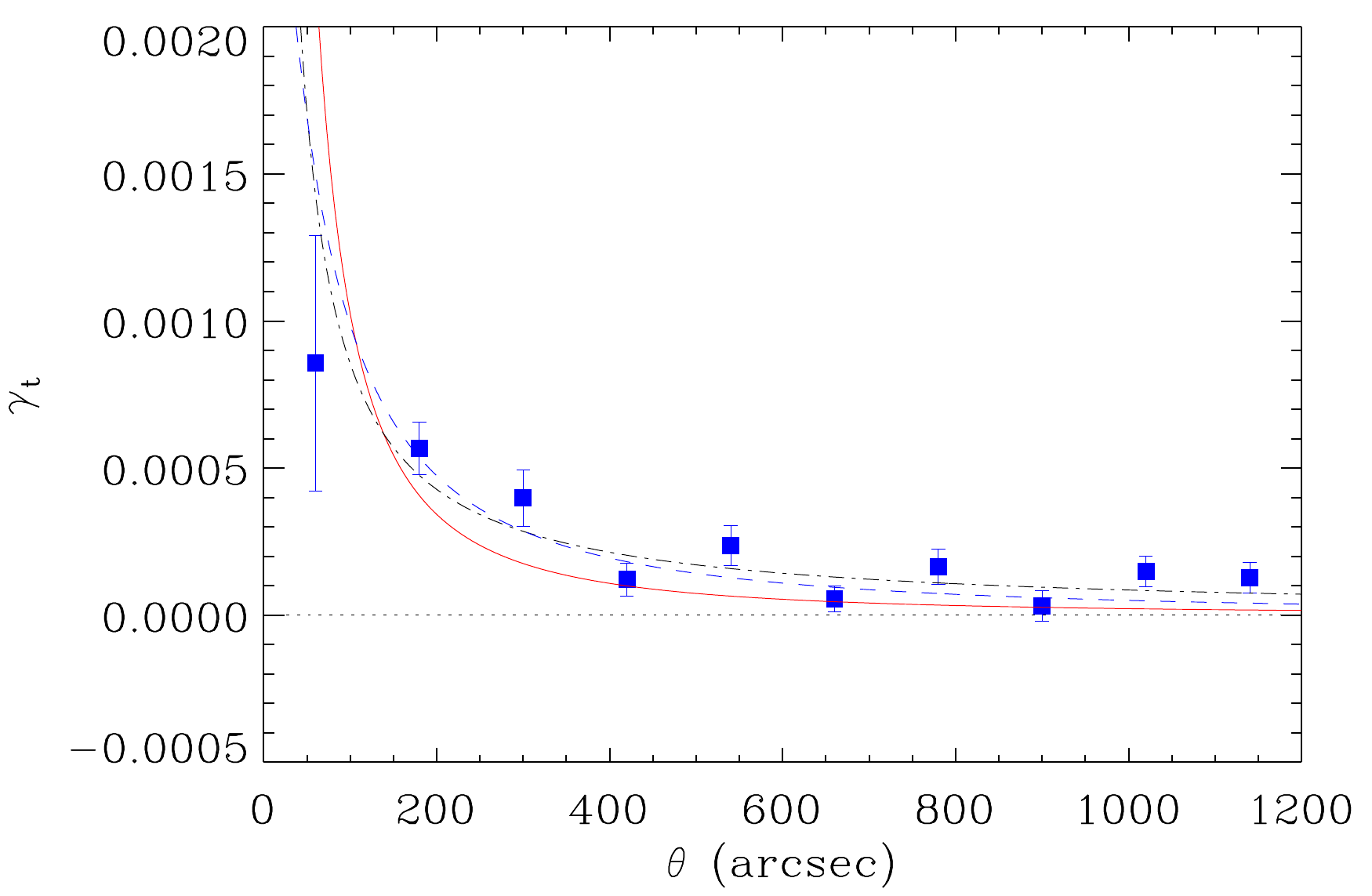}}\hspace{0.5em}
\subfigure{\includegraphics[width=5.5cm]{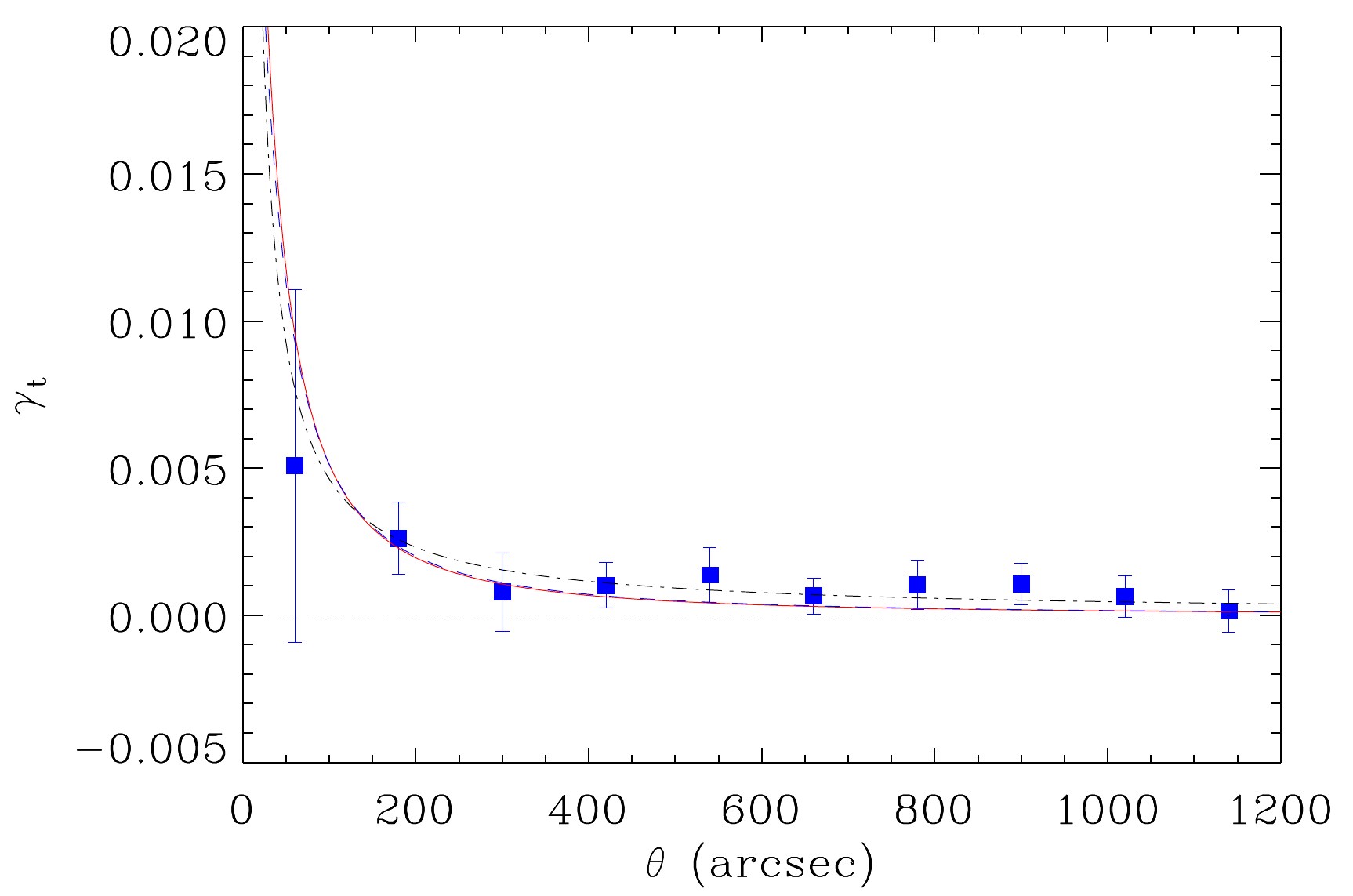}}\hspace{0.5em}
\subfigure{\includegraphics[width=5.5cm]{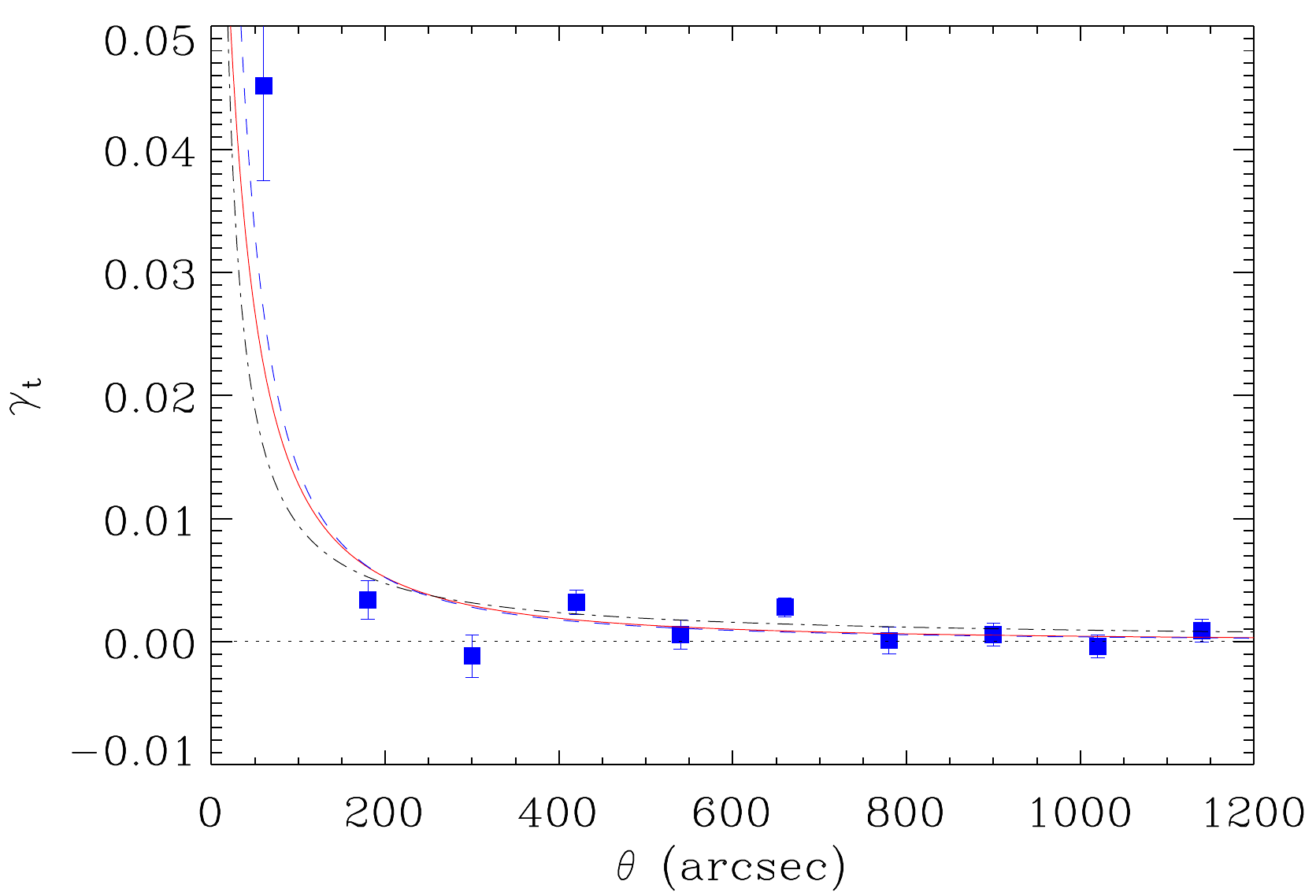}}
\caption{The measured tangential (blue squares) shear for the SDSS
  complete catalogue (\emph{left panel}), the BCG sample (\emph{centre
    panel}) and the SDSS-FIRST matched objects (\emph{right
    panel}). Over-plotted are the best fitting NFW model with a fixed
  c$_f$ (red continuous line), the best-fitting NFW model with a
  variable c$_f$ (blue dashed line) and the best fitting SIS model
  (black dot-dashed line).}
\label{sisnfwmodelsdata}
\end{figure*}

The tangential shear signal measured using the FIRST sources as
background objects and the three lensing samples are shown in
Fig.\,\ref{sisnfwmodelsdata}. Over-plotted are the best-fitting SIS and NFW
models in which the parameters that were used are drawn from the
literature. Also over-plotted is the best-fitting NFW model where the
concentration factor is fitted from the data. The best-fitting
parameters as determined by the data are listed in
Table\,\ref{extractedpar}.

For the SDSS DR10 sample (left panel of Fig.\,\ref{sisnfwmodelsdata})
it is clear that the SIS model is more consistent with the data than
the NFW model in which archival values for the concentration factor
were used. Instead, when the concentration parameter is allowed to
vary, the $\chi^2$ value for the fit decreases by $\sim$25. The
results suggest that the NFW model can be used to constrain the halo
mass of galaxies, but it will predict a mass profile for these sources
that is much shallower than theoretically expected for
$\sim$10$^{12}$\,$\mathrm{M}_{\odot}$ objects. However, both models
appear to predict broadly the same values for the enclosed mass of the
sample $M_{200}$. The extracted values for the SDSS DR10 sample are
consistent (within 1$\sigma$) with results from the Canada France
Hawaii Lensing Survey (CFHTLenS). \citet{velander2013} measured an
Einstein radius and a velocity dispersion for the complete CFHTLenS
lens sample of $\theta_E = 0.136 \pm 0.03$ and $\sigma_u = 97.9 \pm 1.0$
km\,s$^{-1}$ respectively. Also \citet{parker2007} using a
sub-set of that CFHTLenS sample measured an $M_{200}$ value of
$M_{200} = 1.1 \pm 0.2 \mathrm{M}_{\odot}$.

Both the SIS and the NFW profile, in which the concentration factor
was drawn from the literature, are in good agreement with the shear
signal measured from the BCGs (central panel of
Fig.\,\ref{sisnfwmodelsdata}). Additionally, the best fit value for
the concentration factor is roughly the same as the one drawn from
\citet{bullock2001}. This is expected as the NFW profile is primarily
used to predict the matter halos of galaxy clusters in which the dark
matter is the predominant component. The extracted values for the BCG
population are in good agreement with the Hubble Space Telescope (HST)
STAGES study of the Abell 901/902 supercluster where cluster masses in
the range $\left(3.5 < M < 6.5\right) \times10^{13}$ M$_{\odot}$ h$^{-1}$ were
measured \citep{heymans2008}.

Using parameter values from the literature neither the SIS nor the NFW
best fit models can accurately predict the tangential lensing signal
that we have measured around the SDSS-FIRST matched object lens
sample. When allowing the concentration factor to vary, the NFW model
fits to a better degree the data ($\Delta \chi ^2 = 2$) for the
maximum allowed value for the parameter $c_f=12$. The predicted mass
for these galaxies $M_{200}$ is of the order $10^{13}
\mathrm{M}_{\odot}$. No similar work has been conducted on the
SDSS-FIRST sample. Our findings suggest that galaxies, that are bright
in both the optical and the radio, are embedded in very dense
environments on scales $R \lsim 1$ Mpc ($\theta \lsim 150$ arcsec), a
result that is in good agreement with a number of works in the
literature \citep{balogh2004,blanton2005b,blanton2006}.

Finally, it is illustrative to comment on the types of galaxies that
make up the FIRST-SDSS matched catalogue sample. \citet{masters2010}
have shown that the SDSS catalogue parameter {\sevensize FRACDEV}
(which illustrates the fraction of light that is fitted by a de
Vaucouleurs profile) can be used to separate early and late type
galaxies. This is possible since early type elliptical and spirals
with big bulges are traditionally better characterised by a de
Vaucouleur profile and therefore have {\sevensize FRACDEV} $>
0.5$. Using the {\sevensize FRACDEV} information in the SDSS data we
find that the SDSS DR10 sample contains $\sim60$\% early type
galaxies and $\sim40$\% late type galaxies. Contrary to this, in the SDSS-FIRST
matched objects sample $\sim$85\% of the galaxies are early type 
with only $\sim$15\% late type. \citet{courteau2014} showed by
combining the results of a weak lensing, a strong lensing and a
dynamic analysis study that early type SLACS galaxies in the redshift
range $0.1 < z < 0.8$ have an average mass of $M \simeq
2\times10^{13} M_{\odot}$. Our results for the masses of the
FIRST-SDSS lens sample (which are predominantly early-type galaxies)
are in good agreement with the findings of this study (within
2$\sigma$).

\begin{table*}
\caption{Best-fitting parameters for the NFW and SIS dark matter halo
  models, as constrained by the tangential shear measurements around
  the thress lens samples.}
\begin{center}
\begin{tabular}{l|l|c|c|c|c|c|c}
\hline \multicolumn{2}{c|}{} & \multicolumn{2}{|c|}{SDSS DR10} &
\multicolumn{2}{|c|}{BCGs} & \multicolumn{2}{|c}{SDSS-FIRST} \\ \hline
NFW & $c_f$ fixed/fitted by the data & 10 & 1 & 7 & 6 & 7 & 12 \\ &
$\chi ^2$ & 40 & 15 & 5.2 & 5.0 & 27 & 25 \\ & $r_{200} [Mpc]$ &
0.31$\pm$0.04 & 0.22$\pm$0.03 & 0.51$\pm$0.19 & 0.50$\pm$0.18 &
0.90$\pm$0.16 & 0.90$\pm$0.16 \\ & $M_{200}$ [$\times
  10^{12}$$M_{\odot}$] & 3.2$\pm$1.3 & 1.2$\pm$0.4 & 15$\pm$14 &
14$\pm$13 & 79$\pm$43 & 80$\pm$42 \\ \hline SIS & $\chi^2$ &
\multicolumn{2}{|c|}{15.5} & \multicolumn{2}{|c|}{4.5}&
\multicolumn{2}{|c}{31} \\ & $\theta_E$ [arcsec]&
\multicolumn{2}{|c|}{0.17$\pm$0.02} &
\multicolumn{2}{|c|}{0.92$\pm$0.25}& \multicolumn{2}{|c}{1.89$\pm$0.32
} \\ & $\sigma_u$ [KM/h]& \multicolumn{2}{|c|}{102$\pm$33 } &
\multicolumn{2}{|c|}{ 294$\pm$154} & \multicolumn{2}{|c}{339$\pm$140}
\\ & $M_{200}$ [$\times 10^{12}$$M_{\odot}$] &
\multicolumn{2}{|c|}{1.3$\pm$0.2} & \multicolumn{2}{|c|}{32$\pm$13} &
\multicolumn{2}{|c}{48$\pm$15}\\ \hline
\end{tabular}
\end{center}
\label{extractedpar}
\end{table*}

\section{Conclusions}\label{con}
In this study we have performed galaxy-galaxy and galaxy-cluster
lensing analyses, combining information from the ovelapping optical
SDSS and radio VLA-FIRST surveys.

The motivation for this work was to illustrate the advantages of using
radio data in shear studies but also to show the additional benefits
of cross correlating them with information in the optical. The VLA
FIRST shapes were found to contain systematics that, unless accounted
for, will most likely bias a radio-only shear study on angular scales
$\theta \lesssim 200$ arcsec. By cross-correlating the shapes of
selected FIRST sources with the positions of all FIRST sources we have
detected a negative tangential shear which is a function of the
angular separation between the FIRST galaxies and which is
inconsistent with zero at the $>10\sigma$ level. The rotated shear
signal was found to be consistent with zero. Further investigation
revealed that the negative tangential shear resembles the shape of the
VLA beam (or PSF) for observations conducted in snap shot mode,
indicating that the signal is of artificial origins. From the
azimuthally averaged negative tangential shear signal, we constructed
a template of the systematic effect, which we then used to correct the
original FIRST shapes.

Using simulations we showed that the cross correlation SDSS-FIRST
galaxy-galaxy lensing signal is mostly unaffected by this type of
systematic. However, we note that our simulations are not fully
representative of the data as they did not include any correlations
between the postions of the FIRST and SDSS galaxies. Such
correlations exist to some degree in the real data and further
analysis has shown that in the presence of these correlations, the
shape systematis do impact the cross-correlation galaxy-galaxy lensing
measurement, thus motivating the shape correction step that we have
applied.

By cross correlating the positions of the SDSS DR10, BCG and
SDSS-FIRST matched objects with the shapes of the FIRST selected
sources, we measure a tangential shear signal that is inconsistent
with zero at the $\sim10\sigma$, $\sim3.8\sigma$ and $\sim9\sigma$
level respectively. At the same time in all three cases the detection
significance of the rotated shear is much lower. The results are
further assessed using a set of measurements designed to reveal any
residual systematics in the data. These tests do not show any obvious
leftover contamination in the data.

The shape of the measured tangential shear using the SDSS DR10 and
FIRST sources on scales $\theta \gtrsim 200$ arcsec is compared to the
signal predicted in a concordance cosmological model, assuming median
redshifts for the SDSS DR10 and FIRST populations of
$z^{\mathrm{SDSS}}_m$=0.53 and $z^{\mathrm{FIRST}}_m$=1.2
respectively. Our measurements agree within 2$\sigma$ with the theoretical
predictions.

We also fitted our galaxy-galaxy lensing measurements using both NFW
and SIS halo models. In doing so we found that both the SIS and the
NFW profiles fit equally well the BCG-FIRST tangential shear
signal. The concentration factor parameter in the NFW model that was
extracted from the literature and the one fitted using our
measurements are in good agreement. The tangential shear from the SDSS
DR10 sample can be fitted relatively well using an SIS profile. The
NFW profile can also fit the data at the same level but only if one
allows the concentration factor to vary. However, the best fitting NFW
profile is much shallower than the values quoted in the
literature. The lensing signal measured around the SDSS-FIRST matched
objects is best fitted with an NFW profile in which the concentration
factor is much greater than what is typically quoted in the literature.

The best-fitting Virial mass $M_{200}$, Einstein radius
$\theta_{\mathrm{E}}$ and velocity dispersion $\sigma_u$ for the SDSS
DR10 and BCG samples agree within 1$\sigma$ with the values quoted in
the literature. However, we find that the measured ensemble mass of
the SDSS-FIRST matched galaxies is $\sim$2 orders of magnitude greater
than the one found for the SDSS DR10 sample. Using the {\sevensize
  FRACDEV} parameter in the SDSS catalogue we find that $\sim$85\% of
the SDSS-FIRST objects are early type galaxies while only $\sim$60\%
of the objects in SDSS DR10 belong to the same
category. \citet{courteau2014} has also showed that early type SLACS
galaxies in the redshift range $0.1 < z < 0.8$ have an average mass of
$M \simeq 2\times10^{13} M_{\odot}$, a result that agrees with our
findings at the 2$\sigma$ level. Our study has therefore shown that
galaxies which are bright in both the optical and radio wavebands are
typically embedded in very dense environments on angular scales of $R
\lsim 1$ Mpc.

\section*{Acknowledgments}
The authors were supported by an ERC Starting Grant (grant
no. 280127). MLB also acknowledges the support of a STFC
Advanced/Halliday fellowship (grant number ST/I005129/1). 

\bibliographystyle{mn2e_plus_arxiv}
\bibliography{shear}

\label{lastpage}
\end{document}